\documentclass[12pt,preprint]{aastex}

\shorttitle{Yellow Supergiants in M31}

\shortauthors{Drout et al.}

\begin{document}

\title{Yellow Supergiants in the Andromeda Galaxy (M31)\altaffilmark{1}}

\author{
Maria R. Drout\altaffilmark{2} and Philip Massey\altaffilmark{3}}
\affil{Lowell Observatory, 1400 W Mars Hill Road, Flagstaff, AZ 86001;
maria-drout@uiowa.edu; Phil.Massey@lowell.edu}

\author{Georges Meynet}
\affil{Geneva University, Geneva Observatory, CH-1290 Versoix, Switzerland; georges.meynet@unige.ch}

\author{Susan Tokarz and Nelson Caldwell }
\affil{Smithsonian Astrophysical Observatory, 60 Garden Street, Cambridge, MA 02138;
tokarz@cfa.harvard.edu; caldwell@cfa.harvard.edu}

\altaffiltext{1}{Observations reported here were obtained
at the MMT Observatory, a joint facility of the University of Arizona and the Smithsonian Institution.}

\altaffiltext{2}{Research Experience for Undergraduates (REU) participant, Summer 2008. Present address:
Department of Physics \& Astronomy, University of Iowa, Iowa City, IA 52245.}

\altaffiltext{3}{Visiting Astronomer, Kitt Peak National Observatory, National Optical Astronomy
Observatory, which is operated by the Association of Universities for Research in Astronomy, Inc.
(AURA) under cooperative agreement with the National Science Foundation.}

\begin{abstract}

The yellow supergiant content of nearby galaxies can provide a critical test of stellar evolution theory, bridging the gap
 between the hot, massive stars and the cool red supergiants.  But, this region of the color-magnitude diagram is dominated by foreground
contamination, requiring membership to somehow be determined. Fortunately, the large negative systemic velocity of M31, coupled to its high rotation rate, provides the means for separating the contaminating
foreground dwarfs from the {\it bona fide} yellow supergiants within M31. We obtained radial velocities
of $\sim$2900 individual targets within the correct color-magnitude range corresponding to masses of 
12$M_\odot$ and higher.  A comparison of these velocities to those  
expected from M31's rotation curve reveals 54 rank 1 (near certain) and 66 
rank 2 (probable) yellow supergiant members, indicating a foreground contamination $\geq$ 96\%.
 We expect some modest contamination from Milky Way halo giants among the remainder, particularly for the rank 2 candidates,
 and indeed follow-up spectroscopy of a small sample
 eliminates 4 rank 2's while confirming 5 others. We  find
  excellent agreement between the location of yellow supergiants in the H-R diagram and that predicted
 by the latest Geneva evolutionary tracks which include rotation. 
 However, the relative number of yellow supergiants seen as a function of mass varies from that predicted
 by the models by a factor of $>$10, in the sense that more high mass yellow supergiants are predicted than are actually observed.   Comparing the total number 
 (16) of $>20M_\odot$ yellow supergiants
 with the estimated number (24,800) of unevolved O stars indicates that
 the duration of the  yellow supergiant phase is $\sim$3000 years.
 This is consistent with what the 12$M_\odot$ and 15$M_\odot$ evolutionary tracks predict, but disagrees with the 20,000-80,000
 year time scales predicted by the models for higher masses.
  \end{abstract}

\section{Introduction}
\label{Sec-intro}

The color magnitude diagrams (CMDs) of nearby galaxies reveal the details of stellar evolutionary processes, if our eyesight
is keen enough.    In recent years both the data and theory have made considerable advances.  On the one hand, large format
CCD cameras have allowed comprehensive photometry of the resolved stellar content of nearby galaxies, such as that of the
Local Group Galaxy Survey (LGGS), which imaged those Local Group galaxies with active star formation (Massey et al.\ 2006, 2007a, 2007b).
On the other hand, recent advances in stellar evolutionary theory have demonstrated the important role that rotation plays in the
evolution of massive stars (see, for example, Maeder \& Meynet 2000; Meynet \& Maeder 2003, 2005).   

\begin{figure}[ht]
\epsscale{0.6}
\plotone{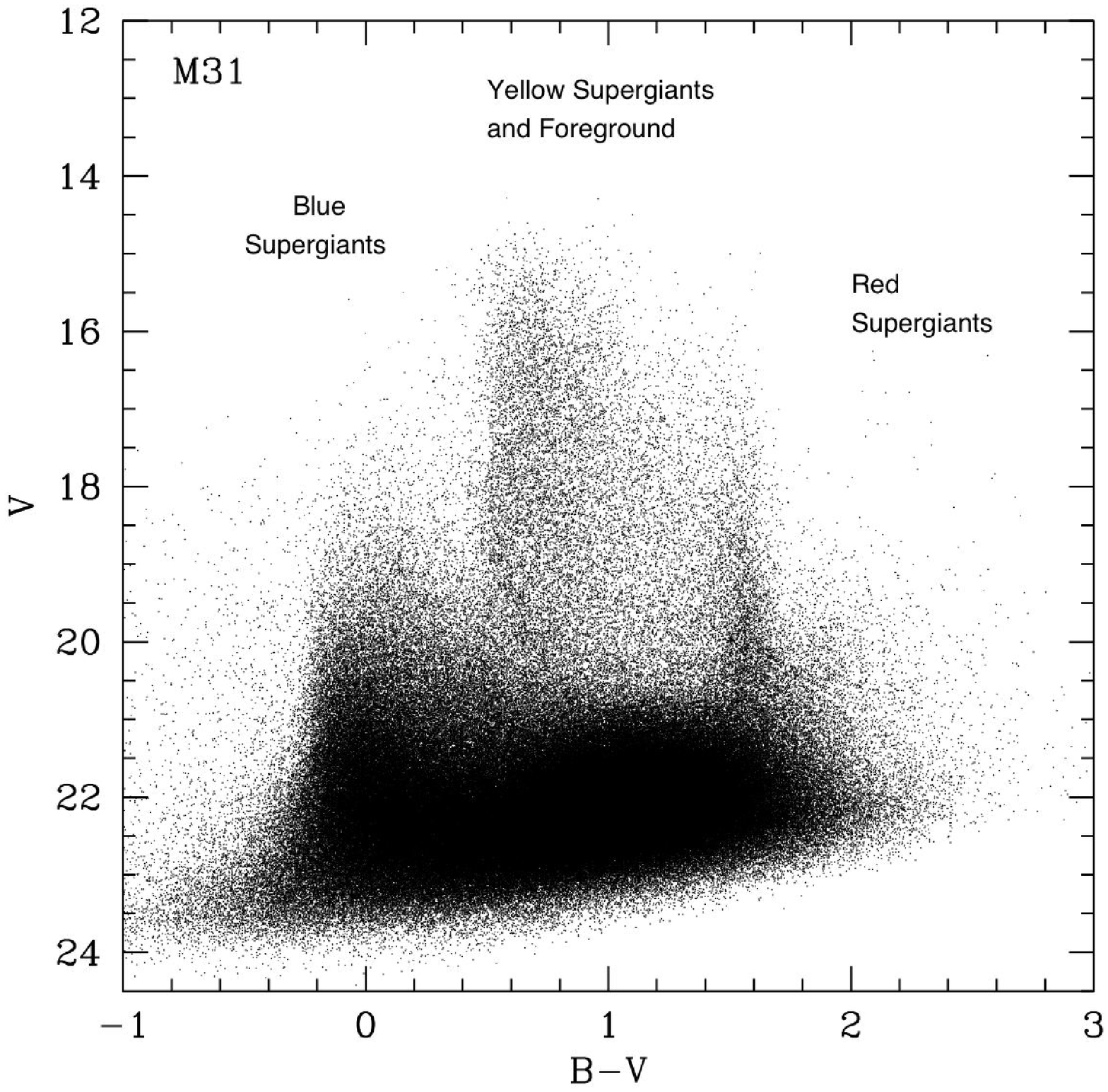}
\plotone{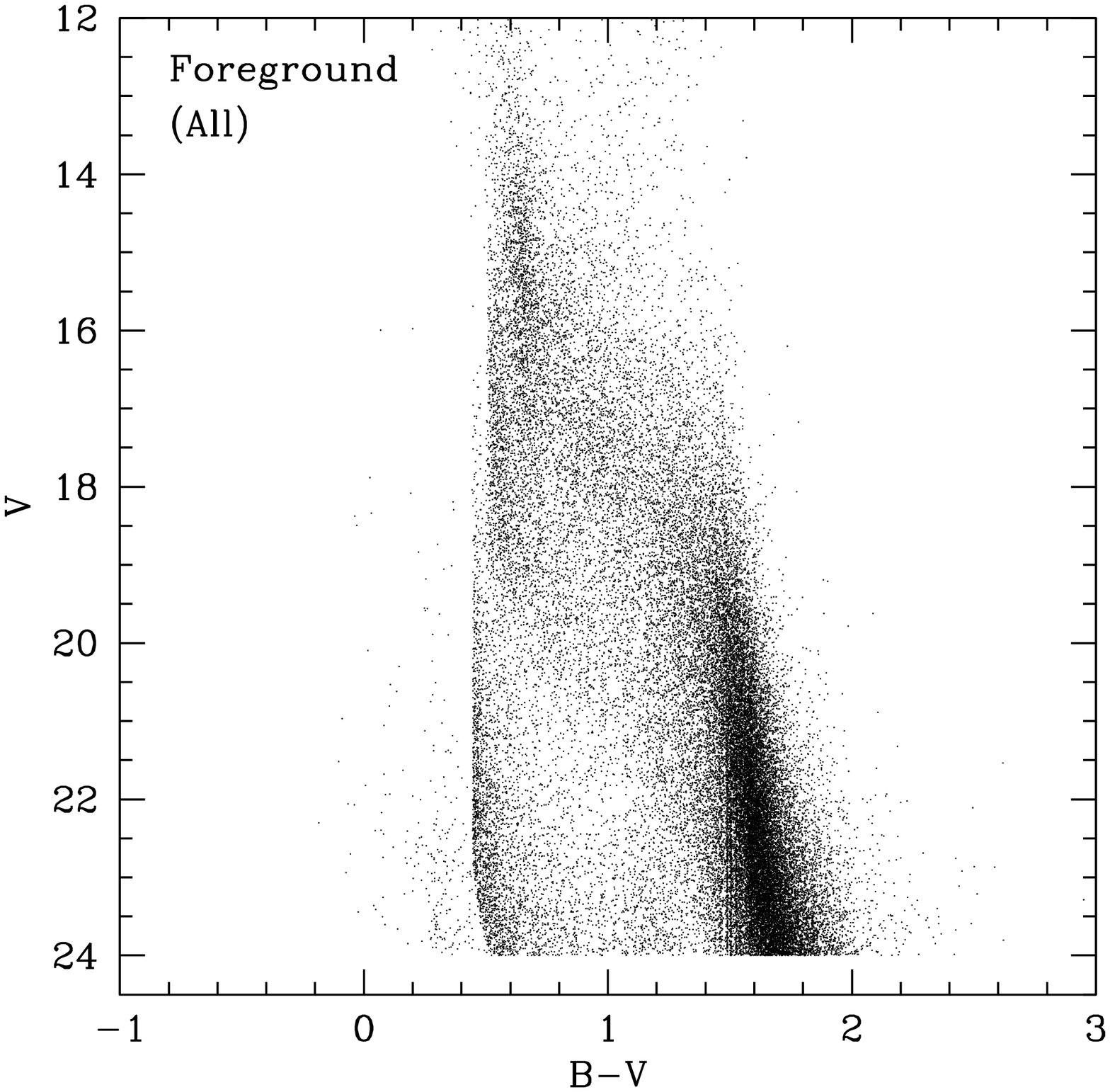}
\caption{\label{fig:CMD} Color-magnitude diagrams. {\it Upper:} The LGGS photometry is used
to construct a color-magnitude diagram for M31.  {\it Lower:}  The regions of large foreground
contamination in the M31 diagram can be inferred from a CMD constructed using the
Besancon model (Robin et al.\ 2003) of the Milky Way.}
\end{figure}

Consider the optical CMD of the Andromeda Galaxy (M31), shown in Figure~\ref{fig:CMD} {\it upper}. The unevolved stars are found on the left, in the section labeled
``Blue Supergiants".   However, this ``blue plume" in such diagrams actually contain a mix of both unevolved
main-sequence and more evolved blue supergiants (Freedman 1988); it also contains a smattering of Wolf-Rayet stars,
the evolved descendants of the most massive O-type stars.   Massey et al.\ (1995) have emphasized that the optically
brightest stars in this region of the CMD are not, in fact, the most bolometrically
luminous or massive---rather, the brightest stars are dominated by B- and A-type supergiants, while more massive (and luminous) O-type
stars are fainter as most of their radiation lies in the far-UV. 

\begin{figure}
\epsscale{0.48}
\plotone{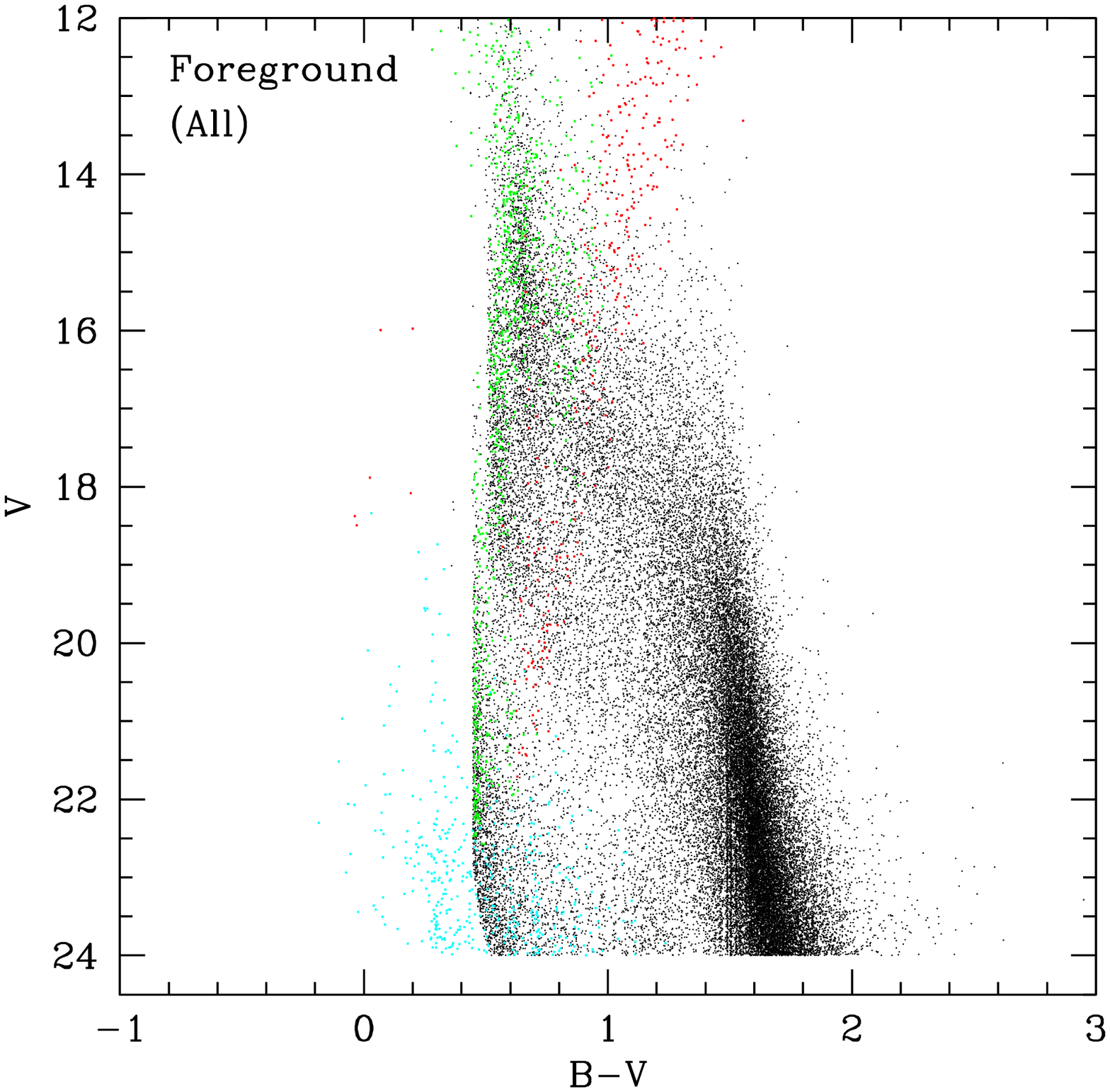}
\plotone{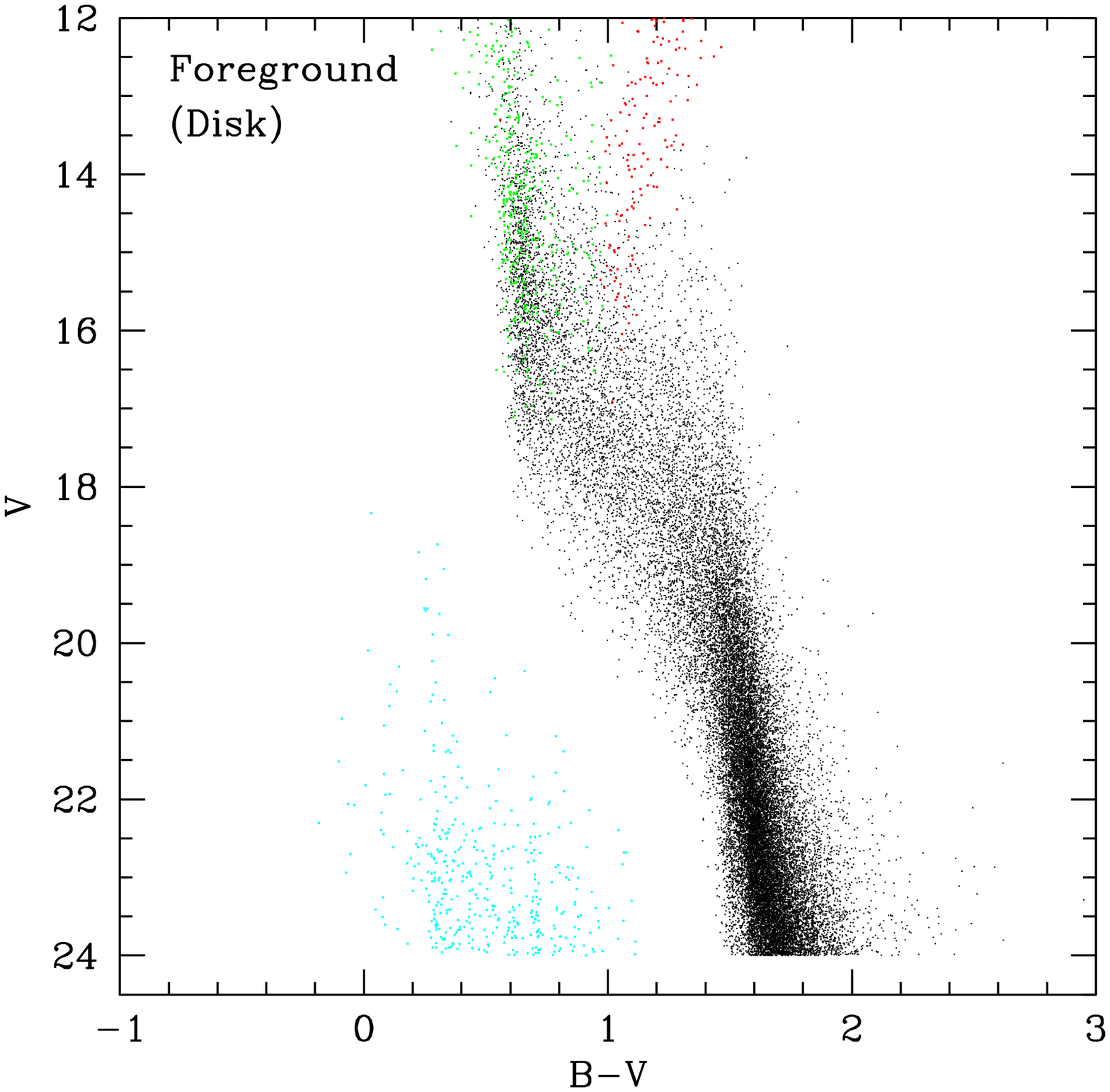}
\plotone{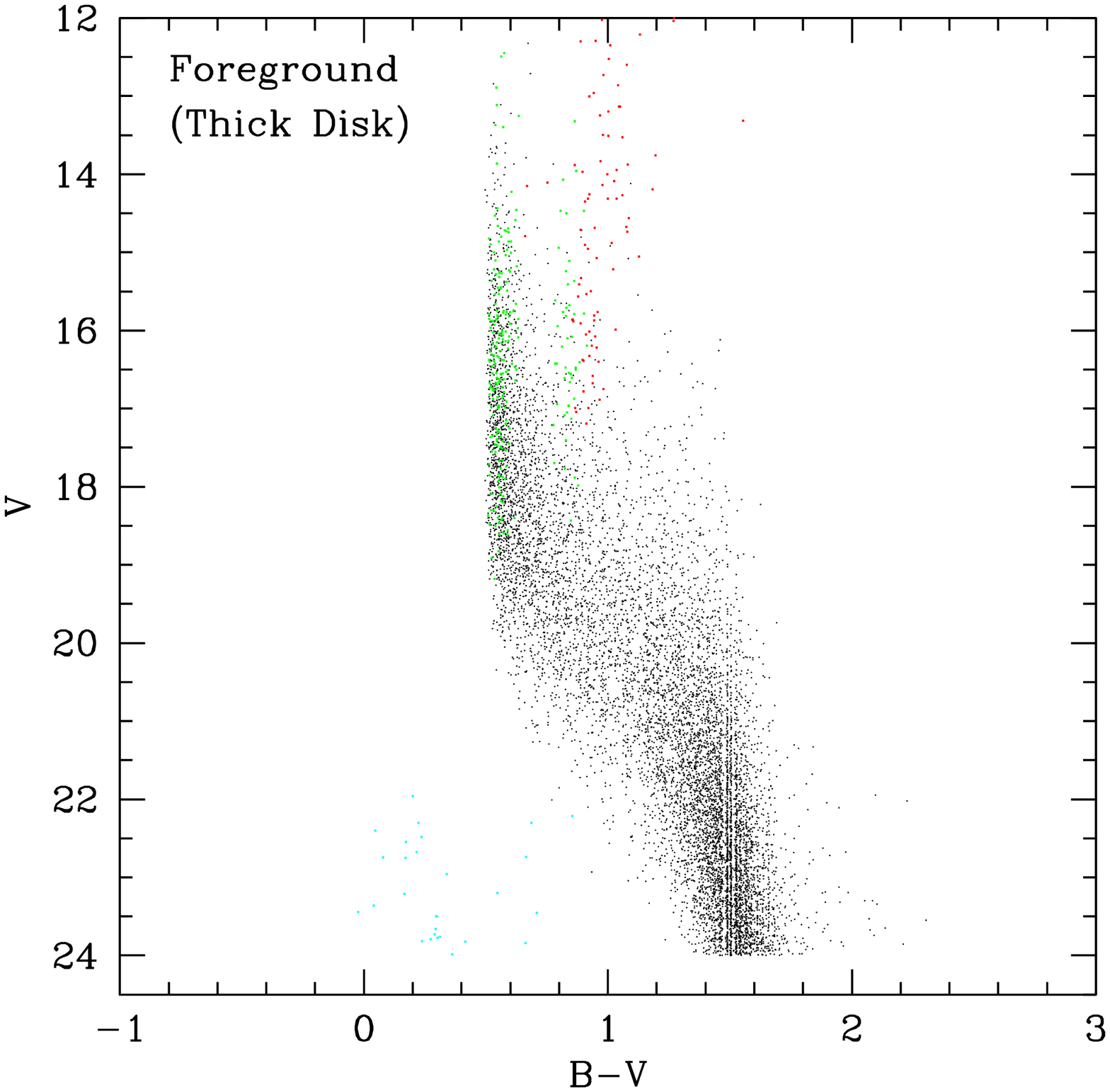}
\plotone{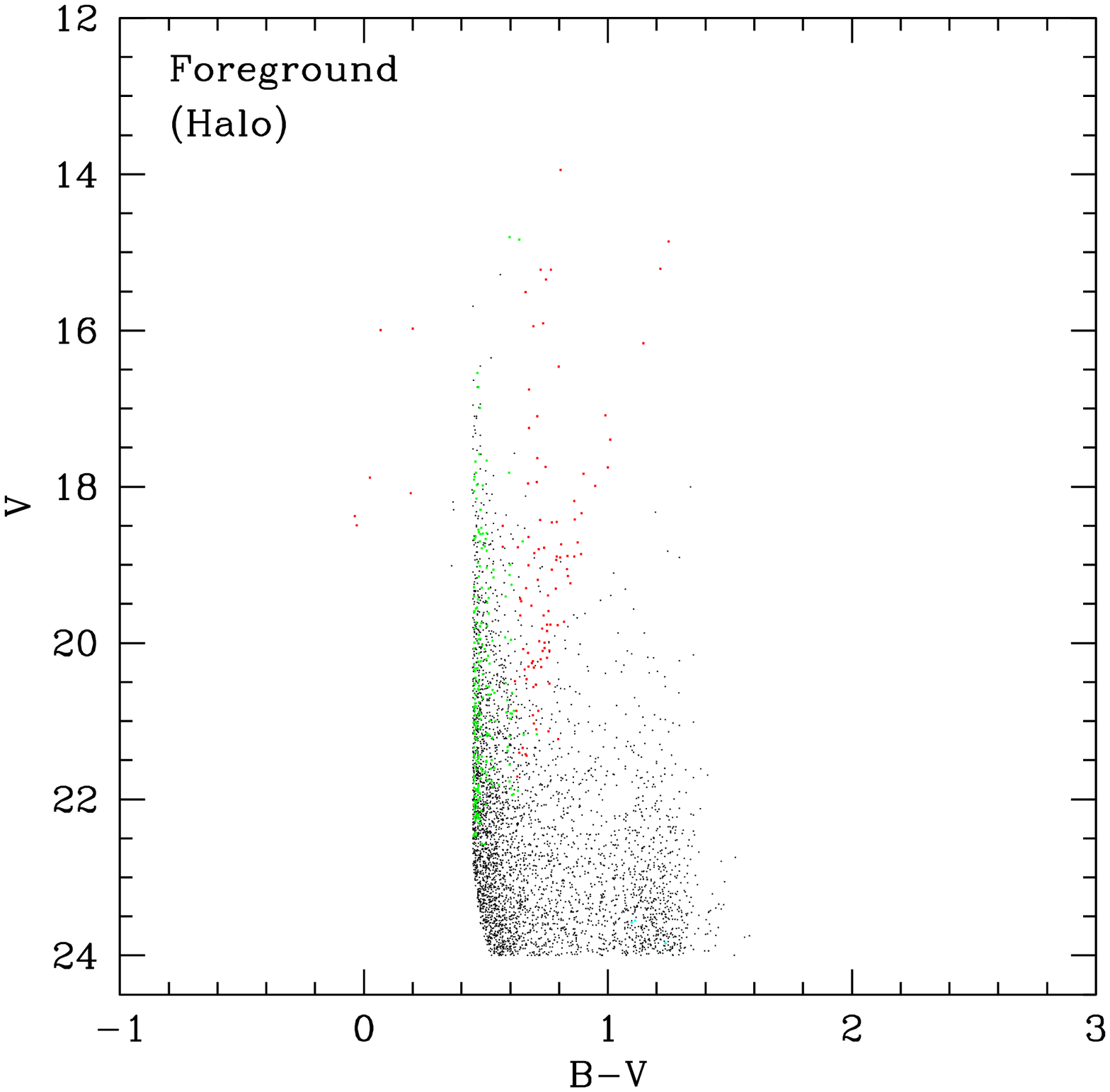}
\caption{\label{fig:CMDFore} Components of the foreground contamination of the M31 field identifed using
the Besancon model.
 In
the upper left we show the expected foreground contamination of the M31 field, where we have
color-coded the giants (red), sub-giants (green),  and the white dwarfs (cyan).  The remaining points are primarily
main-sequence dwarfs.  In the other three panels we further break down the distribution into the disk component
(upper right), the thick disk component (lower left) and the halo component (lower right). }
\end{figure}

The central portion of the CMD contains the yellow supergiants, and at the extreme right, the red supergiants.  But, redwards of the blue plume
caution is advised in our interpretation, as foreground contamination may dominate.    We demonstrate this in Figure~\ref{fig:CMD} {\it lower} by using the Besancon model (Robin et al.\ 2003) of the Milky Way 
to construct a theoretical CMD of the expected foreground contamination, using 
the same galactic coordinates as M31, and covering the same area.  
 For the bright stars ($V<20$) redwards of the blue plume ($B-V>0.4$) the foreground
contamination is $>70$\%.  Note that the features in the M31 ``yellow plume" area are well reproduced
by the Besancon model.  In Figure~\ref{fig:CMDFore} we break down this foreground contamination into
its various components.  The disk populations clearly dominates, but even among the brighter
stars there will be some contamination by halo giants and sub-giants.

Our group is engaged in the long-term process of characterizing the massive star populations of nearby galaxies from one side of the
CMD to the other.   In order to relate this to evolutionary theory, we must succeed in two things: first,
to be able to clean foreground stars from the sample, and secondly to provide a transformation of
observed properties to physical properties.  We have recently undertaken this for the red supergiants of M31 (Massey et al.\ 2009); here we turn our attention to the yellow supergiants.

\subsection{Yellow Supergiants as a Magnifying Glass}
\label{Sec-magglass}
Yellow supergiants (F- and G-type) are extremely rare, as they represent  a very short-lived phase in the evolution of massive stars.
Their numbers and location in the H-R diagram (HRD) are very sensitive to the uncertain values
of the mass-loss rates for massive stars, and how convection and other mixing processes are
treated (Maeder \& Meynet 2000).  As Kippenhahn \& Weigert (1990) put it, ``[The yellow supergiant] phase
is a sort of magnifying glass, revealing relentlessly the faults of calculations of earlier phases."

\begin{figure}
\epsscale{0.48}
\plotone{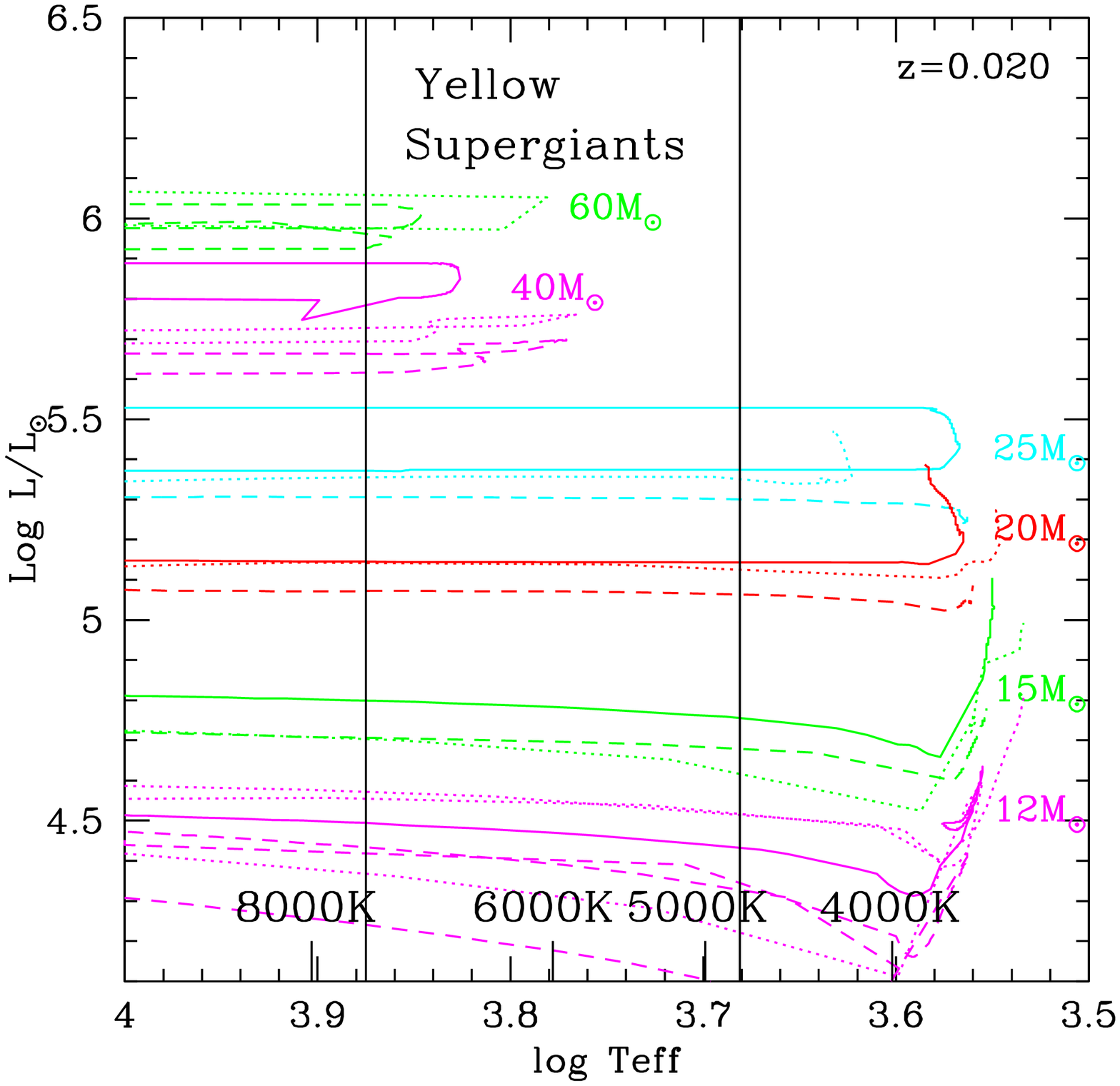}
\plotone{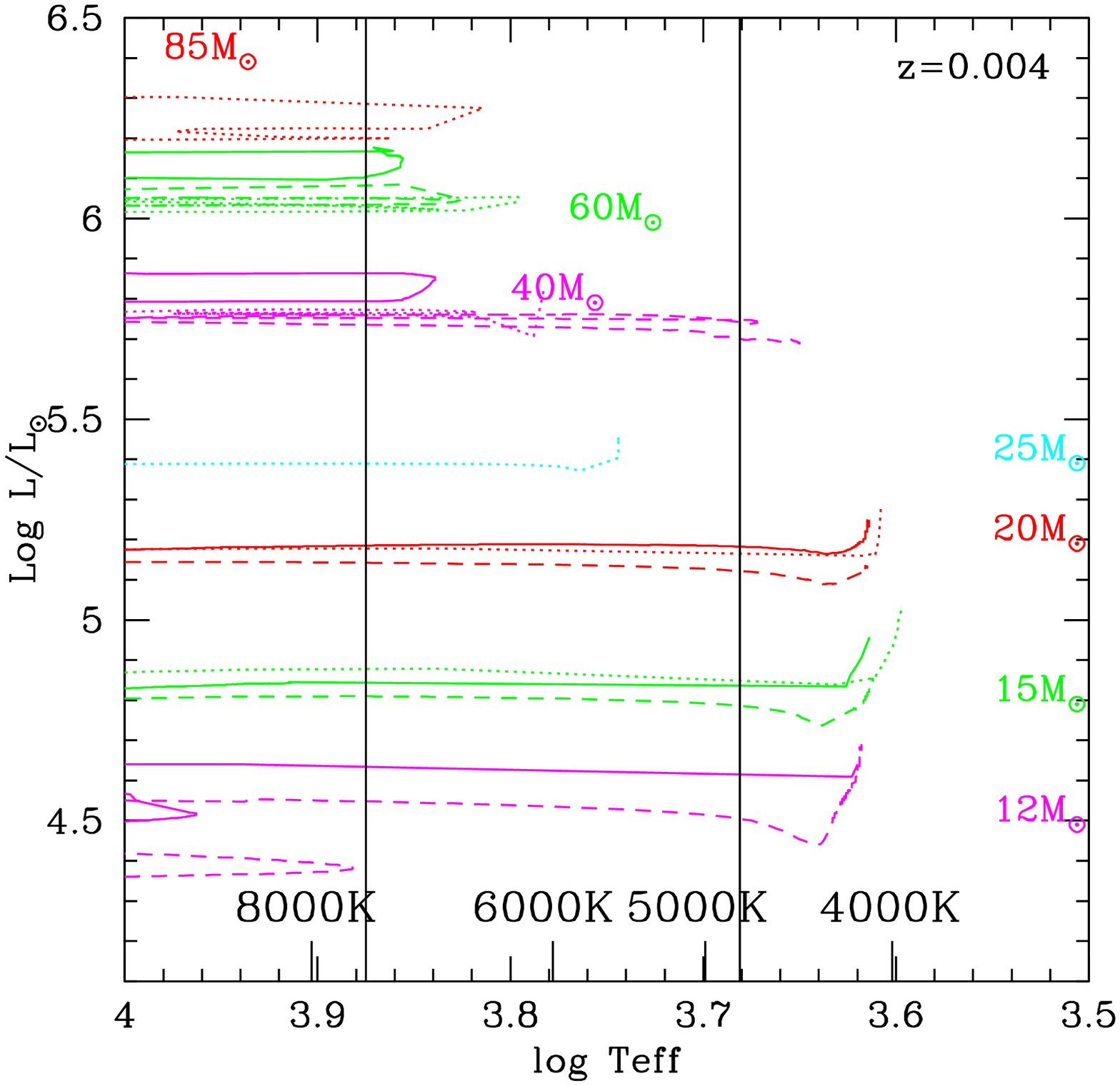}
\plotone{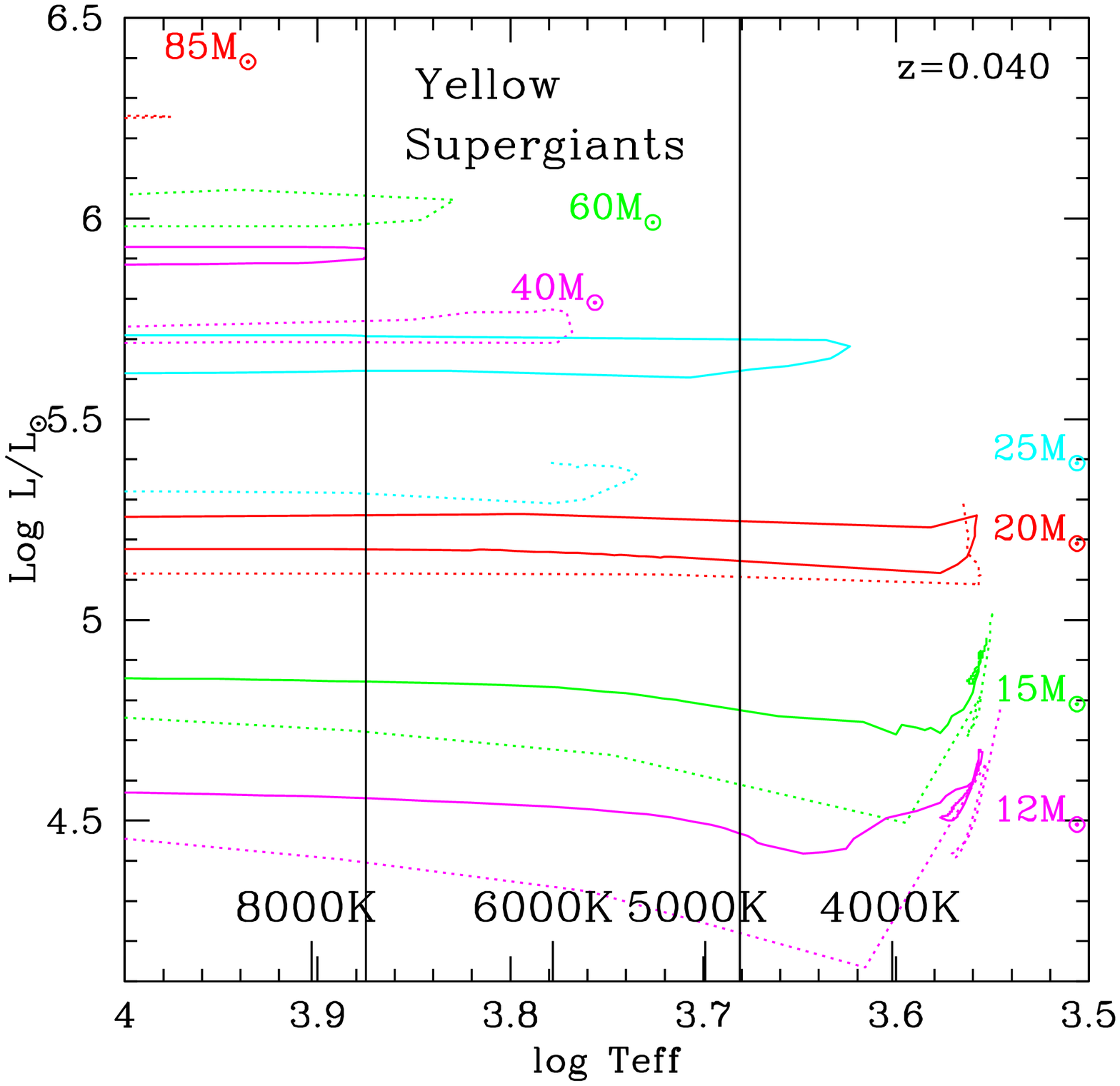}
\caption{\label{fig:models} The Geneva evolutionary tracks.  The solid curves denote the latest models that include an initial rotation of
300 km s$^{-1}$,  the dashed curves are the latest models that include no initial rotation, and the dotted curves are the older, non-rotating
models.  The various versions of the tracks appear in the same color for a given mass to reduce the confusion.  The two vertical lines
denote the yellow supergiant region, taken to be when the models have $4800\le T_{\rm eff}\le 7500$.  The tracks are shown for three
metallicities: $z=0.020$ is characteristic of the solar neighborhood, $z=0.004$ is characteristic of the SMC, and $z=0.040$ is characteristic
of M31.}  
\end{figure}

Exactly how sensitive our expectations should be to the details of the models is illustrated by 
comparing the various Geneva evolutionary models (Schaller et al.\ 1992; 
Charbonnel et al.\ 1993; Schaerer et al.\ 1993; Maeder \& Meynet 2001; 
Meynet \& Maeder 2003, 2005) shown in Figure~\ref{fig:models}
and the corresponding lifetimes for the yellow supergiant stage given in Table~\ref{tab:ages}\footnote{Note that the 20$M_\odot$ and 25$M_\odot$ z=0.040 models of
Meynet \& Maeder (2005) were computed using a numerical simplification which resulted in the tracks turning back to the blue at too high
an effective temperature; here we use the recomputed versions mentioned in Massey et al.\ (2009).
We also include here newly completed $z=0.040$ models for 12$M_\odot$ and 15$M_\odot$ which include the effects of rotation.}.  The metallicities shown span a range of 10, from
sub-solar ($z=0.004$, typical of the SMC) to solar ($z=0.020$) to super-solar ($z=0.040$, typical of M31); see Massey (2002) and references therein.
Solid curves denote the latest models computed with an assumed initial rotation of 300 km s$^{-1}$,the dashed curves are the latest models that include no initial rotation, and the dotted curves are the older, non-rotating models. Since different versions (rotating, newer non-rotating, older non-rotating) of the same mass track
differ in luminosity, we have color coded them for clarity; the corresponding initial mass is shown in the same color.  The two solid vertical
lines denote the yellow supergiant region, which we define as having effective temperatures between 4800~K and 7500~K.

We see that in most cases  the models predict a short pass through the yellow supergiant region as stars evolve
from the OB main-sequence stage (off the plot to the left) over to the red supergiant (RSG) region on the right.  However, in some cases, such as the evolutionary
track that includes rotation (solid curve) for 25$M_\odot$ at Galactic metallicity ($z=0.020$) the star then evolves back to the blue side
of the HRD.  At higher masses (40-60$M_\odot$) the models predict that a star will double back to the blue side while a yellow supergiant.
The lifetimes for the yellow supergiant stage given in Table~\ref{tab:ages} are all very short, typically a few tens of thousands of years.
The main sequence lifetimes of these stars are a few million years, so, in general, the yellow supergiant lifetimes
 are on the order of 1\% or less.  In the extreme cases (lifetimes of the yellow supergiant phase of $\sim 3000$ yr for a 12$M_\odot$ star
with 15~Myr lifetime) it is of order 0.02\%. We do see from Table~\ref{tab:ages}, however, exactly how sensitive these yellow supergiant
lifetimes are to the exact treatments of the models.  At Galactic metallicities, the models computed with an initial rotational speed of 300 km s$^{-1}$
(``S3") indicate a 10$\times$ shorter yellow supergiant phase than models with the same assumptions but with no initial rotation (``S0"). 

There are also clearly differences in the expected {\it locations} of the yellow
supergiants in the HRD, particularly those of the highest luminosities.  According to the models, for instance, at $\log L/L_\odot \sim 5.8$~dex
we would expect to see some yellow supergiants at Galactic ($z=0.020$) metallicities, but only the hotter ones (i.e., $T_{\rm eff}>$6000~K),
while at lower luminosities we should find stars populating throughout the region.  The duration of the yellow supergiant phase is not
significantly shorter for a higher mass (luminosity) star than for a lower mass (luminosity) star, according to Table~\ref{tab:ages}---if
anything the opposite is true---so simply determining the upper luminosity limit to yellow supergiants at various metallicities would be of great
interest for comparison with the models, as well as seeing if the number of higher mass yellow
supergiants really comparable to that of lower masses.    

We emphasize that such a test is new, and avoids some of the selection biases that may dominate
comparisons of one population of massive stars to another, such as comparing the number
of yellow and red supergiants, or the number of red supergiants and Wolf-Rayet stars.
 Although those tests are invaluable, they require a thorough
understanding of the completeness of the surveys of such objects.  We are making progress
towards gathering the information that make such further tests possible (see summary given by Massey 2009).

\subsection{Separating the Wheat from the Chaff}
\label{Sec-wheat}

\begin{figure}
\epsscale{0.55}
\plotone{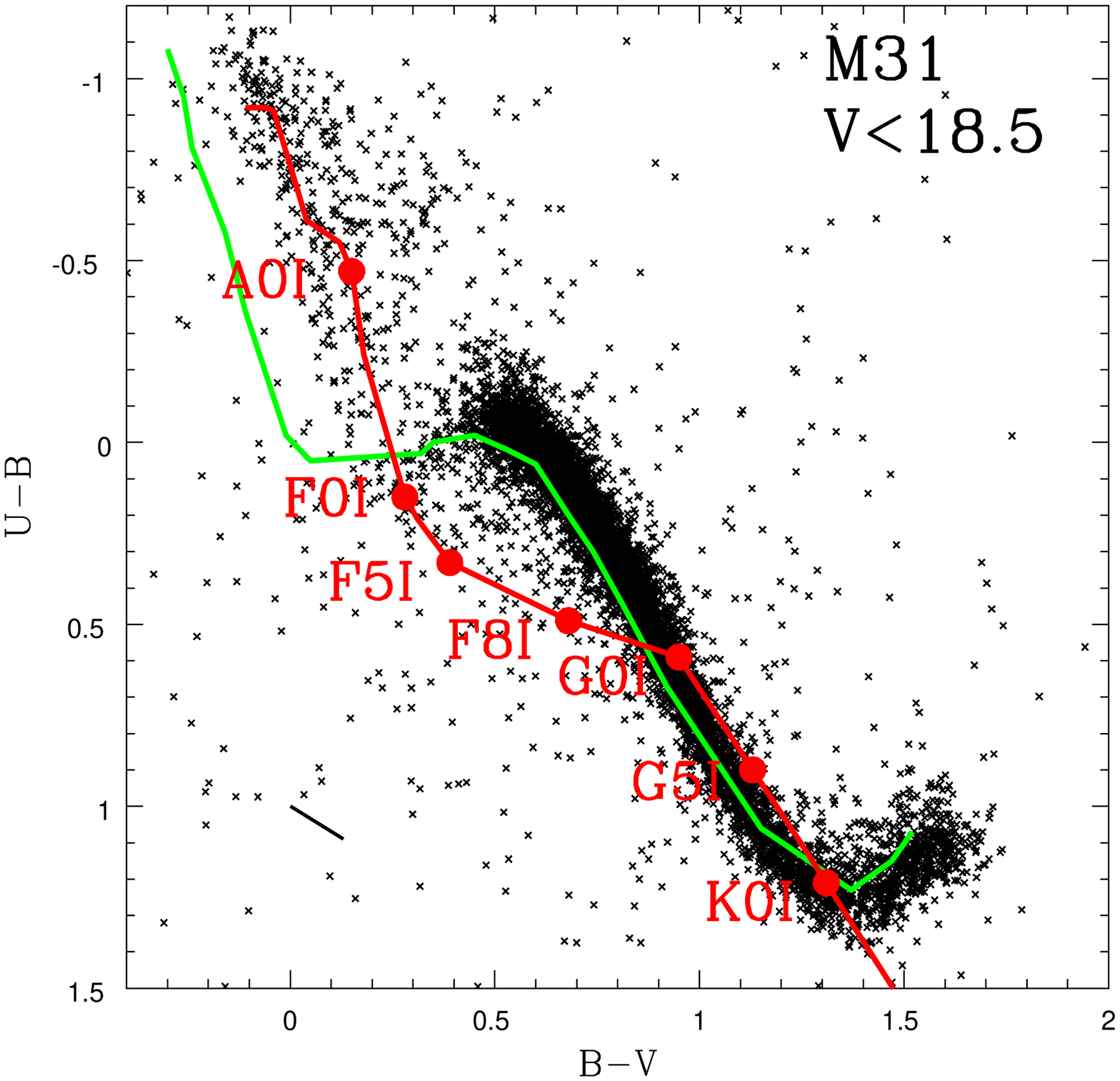}
\plotone{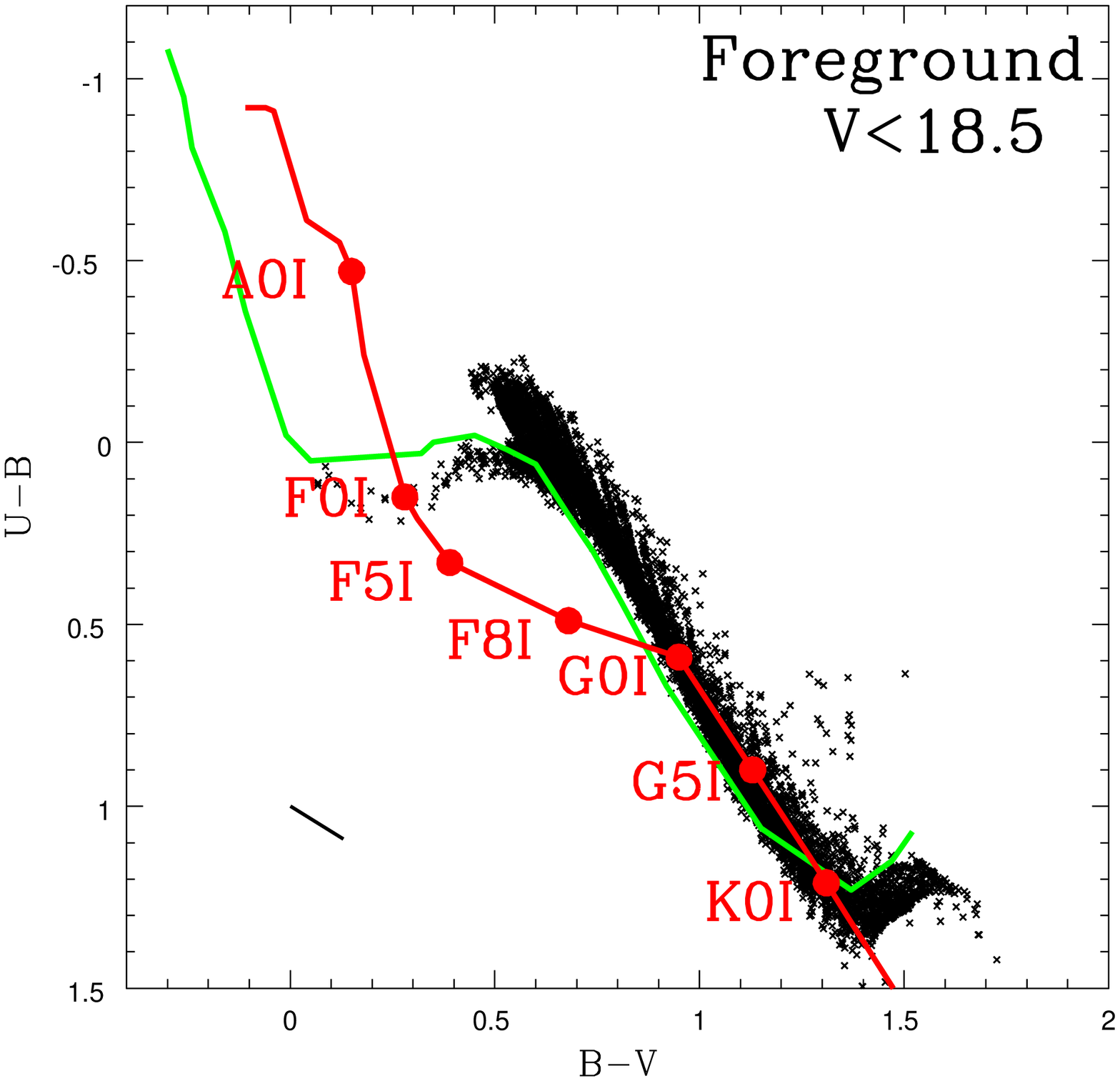}
\caption{ \label{fig:2color} Two-color diagram.  The intrinsic two-color relationships
are shown for dwarfs (green) and supergiants (red), from FitzGerald (1970).  
{\it Upper:} The points show the photometry from the Local Group Galaxy survey of M31
(Massey et al.\ 2006).  The supergiant sequence has been reddened
for a typical $E(B-V)=0.13$, with the reddening vector shown by the short line in
the lower left. {\it Lower:} The points show the approximate foreground contamination from the
Milky Way for the
same solid angle and Galactic latitude and longitude as the M31 photometry. The data come from a simulation with the
Besancon model (Robin et al.\ 2003). }
\end{figure}

We have previously found that it is straightforward to
separate (extragalactic) red supergiants from (foreground) red dwarfs and giants by using a two-color diagram (Massey 1998a,
Massey et al.\ 2009), but this discrimination does not extend to the yellow supergiants.  In Figure~\ref{fig:2color} {\it (upper)} we show the intrinsic color lines from
FitzGerald (1970).  The dwarf sequence is shown in green, and the supergiant sequence in red.  We have reddened the supergiant sequence
slightly, by $E(B-V)=0.13$, corresponding to the typical reddening of a massive star in M31 (Massey et al.\ 2007a); this slight adjustment down
and to the right is shown by the short bar in lower left.  Superimposed on this two-color diagram are the stars brighter than $V=18.5$ from
the Local Group Galaxies Survey (LGGS) photometry given by Massey et al.\ (2006) for stars in M31.

For the mid F-type stars, there is a separation in the sense that for a given $B-V$ a supergiant will have a larger $U-B$.  For early A-type
supergiants this trend is reversed; i.e., supergiants will have a more negative $U-B$.  However, as is clear from this plot there is little or no
separation in the two-color diagram for supergiants of late A or early F, or for supergiants of late F through early K. 
Examination of other colors (using model atmospheres) failed to identify any more suitable diagnostic two-color tool.

It is also clear from distribution of points in this figure that the majority of blue stars are expected to be supergiants, while the majority of stars
of later types may well be foreground, as we argued above.   We illustrate this further in Figure~\ref{fig:2color} {\it (lower)}, where we have used the Besancon simulation
of the Milky Way (Robin et al.\ 2003) to show the expected distribution of foreground stars in such a diagram. The comparison between the two sides of the figures reveals where we expect foreground
stars to dominate.

Given the lack of two-color discriminants, what then are our options to identify a relatively complete sample of yellow supergiants that can be used to test the stellar evolution models?
 Uncertain distances and reddening complicate the analysis of a galactic sample.
 In her compilation of 949 supergiants known in the Milky Way, Humphreys (1978) lists just 21 supergiants of spectral type F0-G9 I 
(i.e., about 2\%).  Even so, several of these have dubious cluster memberships and, hence, uncertain luminosities.  In the 
Magellanic Clouds there are less than a dozen spectroscopically confirmed F and G supergiants (Oestreicher \& 
Schmidt-Kaler 1999), although this deficiency may be largely due to the lack of adequate spectroscopic studies in the correct 
magnitude/color range, a situation we ourselves hope to remedy in the not too distant future.  

We have concluded that M31 provides the best laboratory for conducting such tests at present, as its kinematics allow us to overcome
the problems posed by foreground contamination of its CMD.  M31 possesses a large
 negative systemic velocity ($\sim -300$ km s$^{-1}$) and a high rotation rate (240 km s$^{-1}$), making it relatively
 straightforward to demonstrate membership based on radial velocities.  Gilbert et al.\ (2006) and Koch et al.\ (2008) similarly
 used radial velocities to separate M31's red giant members from foreground contamination. 
 The recent LGGS photometry of M31 (Massey et al.\ 2006) provides the means for selecting candidate stars for radial velocities,
 and for transforming intrisic colors to physical properties.
 
 In this paper we conduct a census of yellow supergiants in M31, establishing membership, determining physical properties, and making comparisons
 with the current generation of evolutionary tracks. In \S~\ref{OS} we describe our data and 
reduction.  In \S~\ref{A} we illustrate the process by which we separated foreground dwarfs from M31's yellow
 supergiants, and in \S~\ref{HR} we present the comparison of our results to the current evolutionary tracks.  We provide discussion
 and lay out our thoughts for future work in \S~\ref{fut}.

\section{\label{OS} Observations and Reductions}

In order to separate yellow foreground dwarfs from the  yellow supergiants,
we used the Hectospec 300 fiber spectrograph on the 6.5-m MMT telescope to obtain radial velocities for $\sim 2900$ stars.  In this section we describe the sample selection, data acquisition,
and reductions.

\subsection{\label{SS} Sample Selection}

Our sample of objects was selected to have $V<18.5$ (roughly corresponding to $\log L/L_\odot\sim 4.4$)
to provide adequate signal-to-noise
for our spectroscopy.  The color range was originally restricted to $U-B>-0.4$ with
$0.4\leq B-V \leq 1.4$ in accordance with the range of $B-V$ for which dwarfs and supergiants
 cannot be distinguished photometrically; however, we found that relaxing the color range to
$0.0\leq B-V \leq 1.4$ added only a small percentage of additional stars.

Our other selection criteria was based upon the need to be able to distinguish the
M31 yellow supergiants from foreground disk dwarfs based upon their radial
velocities.  Inspection of the atlas of Galactic neutral hydrogen by Hartmann \& Burton
(1997) shows that M31 (at $l=121.2^\circ$, $b=-21.6^\circ$) clearly stands out
from the Galactic clutter at a radial velocity of -150  km s$^{-1}$ (see Hartmann \& Burton 1997, p. 87),
 but is somewhat confused by -100 km s$^{-1}$ (pp. 95, 97).  We therefore decided to restrict our 
observations to those areas in M31 where the radial velocities should be $\leq$ -150 km $^{-1}$.  
For computing the expected radial velocity corresponding to a position in M31, we used Rubin
\& Ford (1970), the seminal paper on the rotation of M31.  A least-squares linear fit to the
 radial velocities of the HII regions yields the expected radial velocities Vel$_{\rm expect}$
$${\rm Vel}_{\rm expect} = -295+241.5  (X/R)$$
where X is the distance along the semi-major axis, and R is the radial distance
of the object within the plane of M31.  We found that this approximation works well, producing good agreement
with the more complex two dimension velocity field (Sofue \& Kato 1981), and with other recent
approximations (Hurley-Keller et al.\  2004).   The radial velocities of red 
supergiants in M31 also agrees well with this simple relationship (Massey et al.\ 2009).  

\begin{figure}
\epsscale{0.8}
\plotone{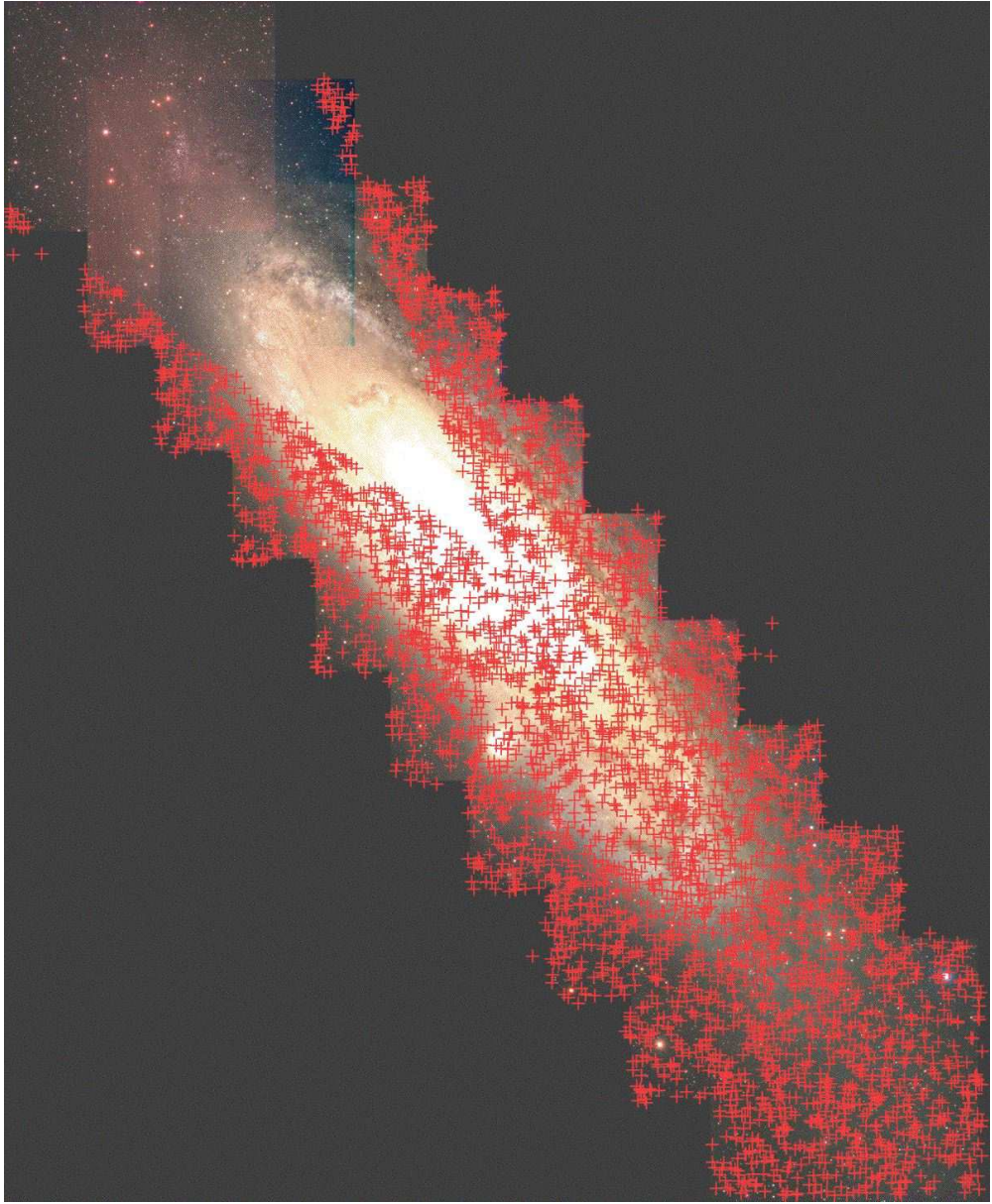}
\caption{\label{fig:sample} The distribution of our sample across the face of M31. The M31 image is a mosaic of the 10
LGGS 36'$\times$36' fields.  The stars selected for our sample are shown by the red points. The requirement that the
expected rotational velocity be $\leq -150$ km s$^{-1}$ results in the ``alligator jaw" pattern.  The figure is slightly smaller
than the area over which the LGGS has photometry, and hence a few points fall outside the image of the galaxy.}
\end{figure}

The result of this selection criterion is that not all of the  area surveyed in the LGGS was included
in our sample: stars along the  south-west half of the semi-major axis of M31 ($X/R\sim -1$) will have radial velocities of $\sim$ $-550$ km s$^{-1}$,
but stars along the north-east half of the semi-major axis ($X/R \sim 1$) will have radial velocities of $-50$ km s$^{-1}$, more positive than our
selection criterion of $-150$ km s$^{-1}$.  
The distribution of stars in our sample is shown in Figure~\ref{fig:sample}, where the ``jaws" in the north-east
(upper left) are due to stars near the semi-major axis having smaller $R$ values then those seen along the edge of the disk.  In all our
sample covered 1.6 deg$^2$ of the 2.2 deg$^2$ of the LGGS survey (i.e., about 73\%).

We were concerned that a few legitimate F or G supergiants {\it might} be too bright
to be included in the LGGS photometry, particularly due to
saturation in the $R$ filter around $R\sim 15.5$.  
We therefore supplemented the LGGS data with 163 bright ($V<16.0$) stars from
the survey of Magnier et al.\  (1992).   In order prevent mixing bright and faint stars
in the same observations, we then divided our overall catalogs into a ``bright" catalog
(349 stars with $V<16.0$) and a ``faint" catalog (3994 stars with $15.5<V<18.5$),
with 61 objects in common.  Due to the constraints of the fiber configurations, not all the objects could be
 observed, but we did manage to observe 68\% of the 4282 catalog targets.

\subsection{Spectroscopy}

Hectospec is a 300 optical fiber fed spectrograph (Fabricant et al.\ 2005) on the MMT
6.5-m telescope.  Observations are obtained in an innovative queue mode where
the observers are the astronomers who were awarded time, but with the observing
program for a given night determined by a queue manager.   The observers are
ably assisted by one of two professional instrument operators, in addition to the
telescope operator. This ``collective" approach spreads out the effects of poor weather throughout
 the observing season, reducing the impact on any one program.  Our observations were
all carried out on eight nights during October 2007 and one night in November 2007.
The 600 line mm$^{-1}$ grating was used, providing a dispersion of 0.55 \AA\ pixel$^{-1}$ and a 
(5 pixel) spectral 
resolution of 2.8 \AA.  The wavelength coverage extended from 4550-7050 \AA.

The fiber configuration files were constructed prior to the observations.  The instrument
has a 1$^\circ$ field of view, a reasonable match to the 3-4$^\circ$ angular extent of
the optical disk of M31 (Hodge 1981) .  Our observations consisted of observing a single
configuration for each of the five fields containing the
brighter stars (``Brt" fields), and multiple configurations of three fields containing the fainter stars (``Fnt" fields),
as listed in Table~\ref{tab:fields}.  Observations each ``Brt" configuration consisted of  3 
consecutive exposures of 10 minutes each, while observations of the ``Fnt" configurations consisted of 3 
consecutive exposures of 15 minutes each.  In the end, we obtained 3116 spectra of 2901 of our catalog targets.

Owing to the logistics of the queue observations, calibration exposures (flat field
and He-Ne-Ar) were taken only in the afternoon, and subsequent to that the grating
might have been tilted to a different angle to accommodate other programs, or even
removed entirely and then replaced for our observations.  We deal with this complication
as described below.

We also needed observations of stars that could serve as templates
for the cross-correlation.  Since the 600 line mm$^{-1}$ is not commonly used,
we obtained our own observations of three radial velocity standards, HD 196850, HD 194071, 
and HD 213014.  HD 213014 was observed on four different nights. 

\subsection{Data Reduction}
The data were all reduced using the ``hectospec" IRAF\footnote{IRAF is distributed by NOAO,
which is operated by AURA under cooperative
agreement with the NSF.  We appreciate the on-going support of IRAF by the
volunteers at the IRAF help ``desk", http://www.iraf.net.} package, designed
specifically for this instrument (Mink et al.\ 2007).
 The data were bias-subtracted, trimmed, and a bad pixel extrapolation
was performed using pre-existing bad pixel maps.  The flat field exposures were
extracted for each fiber and normalized in order to make the pixel-to-pixel corrections.
The He-Ne-Ar arc exposures were then extracted, and used to make a dispersion
solution.  A sixth order chebyshev function was used for this, resulting in 
typically 0.04 \AA\ residuals. 

The program  exposures were extracted, using the dome flat field 
exposures as reference, and wavelength corrected. 
As the He-Ne-Ar exposures were obtained in the afternoon, and the
grating might even be removed and reinserted before the program exposures, a
zero-point shift in wavelength was determined for each of the M31 spectra using the
O I $\lambda 5577$ night sky line.  (As explained below, no such correction could be
made for the very short exposures of the radial velocity standards.)
Consecutive exposures of each M31 configuration were then summed, after cosmic rays
were first identified and removed by comparing a median of the exposures to the
individual exposures.  

For sky subtraction, each fiber had to be first be corrected for its own wavelength dependent throughput, using
either the dome flat exposures, or, preferably, twilight exposures, if the latter had been obtained.
 Each M31 configuration contained both preselected ``clean" sky positions
plus random locations that might prove clean enough to be used as a measure of the sky.
For each program spectrum 6 of these
sky spectra were selected.  These were taken from positions nearby on the array
in order to reduce any scattered light component.  These skys were then used to
construct an average sky for subtraction using the Singular Value Decomposition method
(Mink \& Kutz 2001).  No sky subtraction was needed for the bright radial velocity standards.

\section{\label{A} Analysis}
\subsection{Observed Radial Velocities}

The radial velocities of our 2901 program objects were obtained through a cross-correlation
 with suitable radial velocity standards.  All cross-correlations were performed in the IRAF package
 ``fxcor", which computes radial velocities via Fourier cross-correlation using a Gaussian to find 
the center and width of the calibration peak, following the method of  Tonry \& Davis (1979).
Before the cross-correlations were computed, each 
individual spectrum had its continuum removed by first normalizing and then subtracting 1.0. 
The normalization was done interactively, using a 12th order cubic spline fit to the continuum.  The 
cross-correlations were restricted to the range 4750-5550 \AA\ and 5600-6800 \AA\ in order to avoid the
strongest night-sky emission.

As mentioned above, although the [O I] $\lambda 5577$ emission line was adequate to correct the program 
spectra for the wavelength zero-point (necessitated by the logistics of the queue) there was, ironically,
 no equivalent way to correct the radial velocity ``standards".  In order to resolve this issue, we
 used two velocity templates, created for a related project, generated from 270 line mm$^{-1}$ Hectospec
 observations (Caldwell et al.\ 2009).  The templates were created by first deriving an initial velocity
 from library templates (typically a K giant star) for spectra selected from a catalog of M31 star clusters.
  The spectra were shifted to zero velocity and sorted into crude spectral bins (F- and K-type) at which 
point the best spectra from each type were then combined to make new templates and the whole process repeated.
  When we cross-correlated our ``rv-standards'' with these templates, the velocities produced by the two 
templates varied by $\leq$ 1.1 km s$^{-1}$ for all six observations.  A third template, based upon early-type
 spectra (A-type) gave a velocity $\sim$5 km s$^{-1}$ more positive, and we chose not to use it
 \footnote{We did find that the low-resolution templates produced problems when measured directly against our M31
 observations obtained at higher dispersion, giving inconsistent results on the one field repeated on two nights.  
The differences were strongly correlated with velocities, which we interpreted as being due to cross-correlating 
two spectra of very different dispersions and velocities.}.

In order to obtain suitable velocities for the radial velocity standards the relative velocities of the standards
 to the templates were obtained via cross-correlation and corrected to heliocentric velocities. 
 We were very pleased to find this process yielded velocities that agreed with the IAU adopted velocities to
 better than 5 km s$^{-1}$.  This indicates that if we had simply adopted the standard velocities, with {\it no} corrections,
 the resulting error would not have been large enough to affect our results.

As a final check of our newly adopted velocities, the standards were cross-correlated against each other. 
 HD194071 and the four observations of HD213014 all produced velocities within 0.2 km s$^{-1}$ of the
 standard value value, and within 0.1 km s$^{-1}$ of each other.  HD196850, however, consistently produced velocities 
0.5 km s$^{-1}$ more positive.  Although this value is not significant compared to the expected radial velocity 
difference between M31 yellow supergiants and galactic yellow dwarfs, it still prompted us to reject 
our spectrum of HD196850 as a radial velocity standard template.  We then performed cross correlations of the five spectra of the radial
velocity standards against each of the 3116 spectra of the 2901 program objects. 

There were four objects for which our cross-correlations initially failed:  J004101.24+410434.6, J004129.31+40502.9, J004203.63+405705.8, and J004459.11+412732.7.  The spectra of all four
 objects were examined and strong nebular emission in the region of H${\alpha}$ was evident.  When this region was excluded 
good correlations were found for J004101.24+410434.6, and J004129.31+40502.9, although their associated
 error was about twice as large as the average value of the other program objects.  

\begin{figure}
\epsscale{0.35}
\plotone{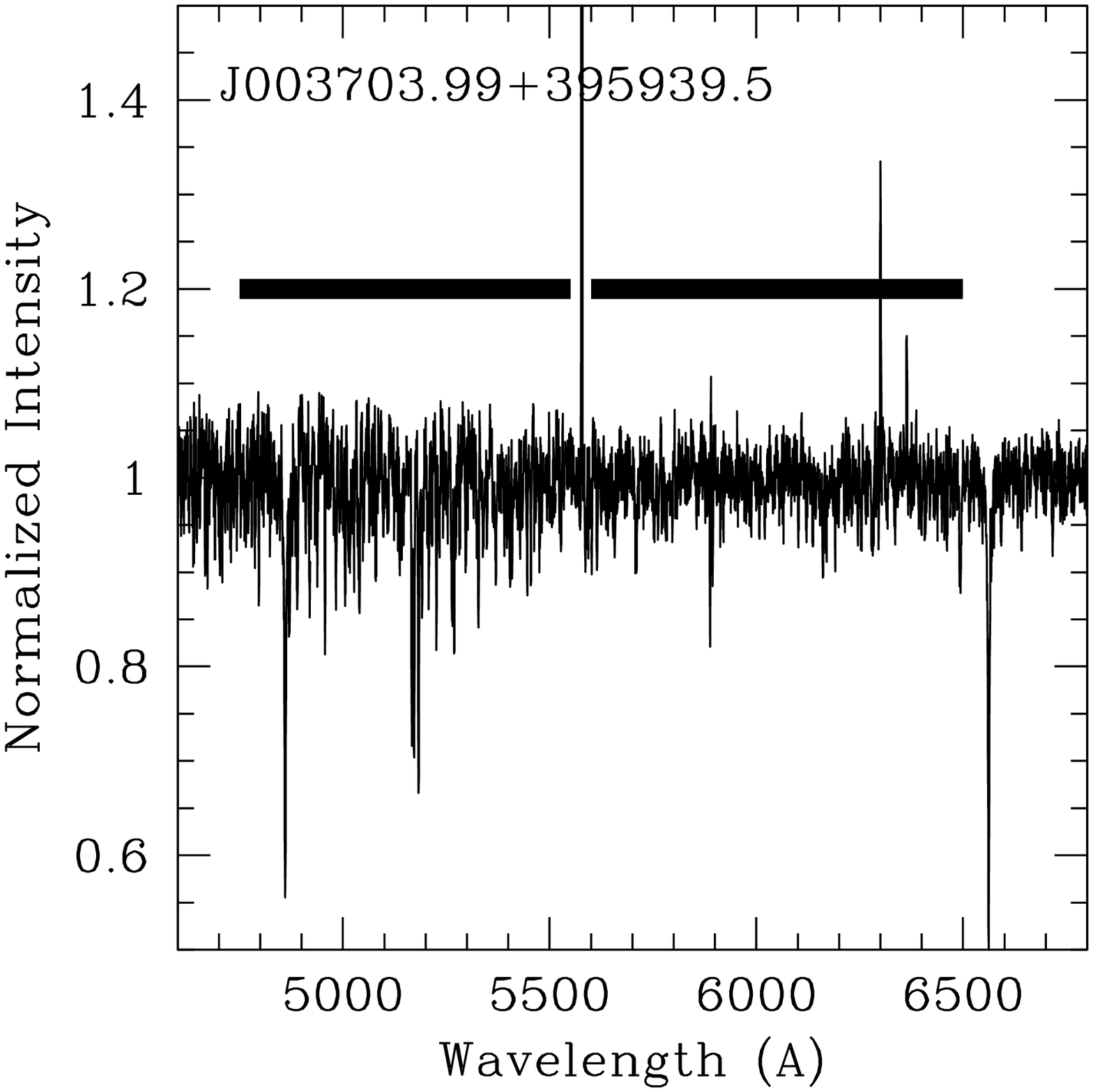}
\epsscale{0.55}
\plotone{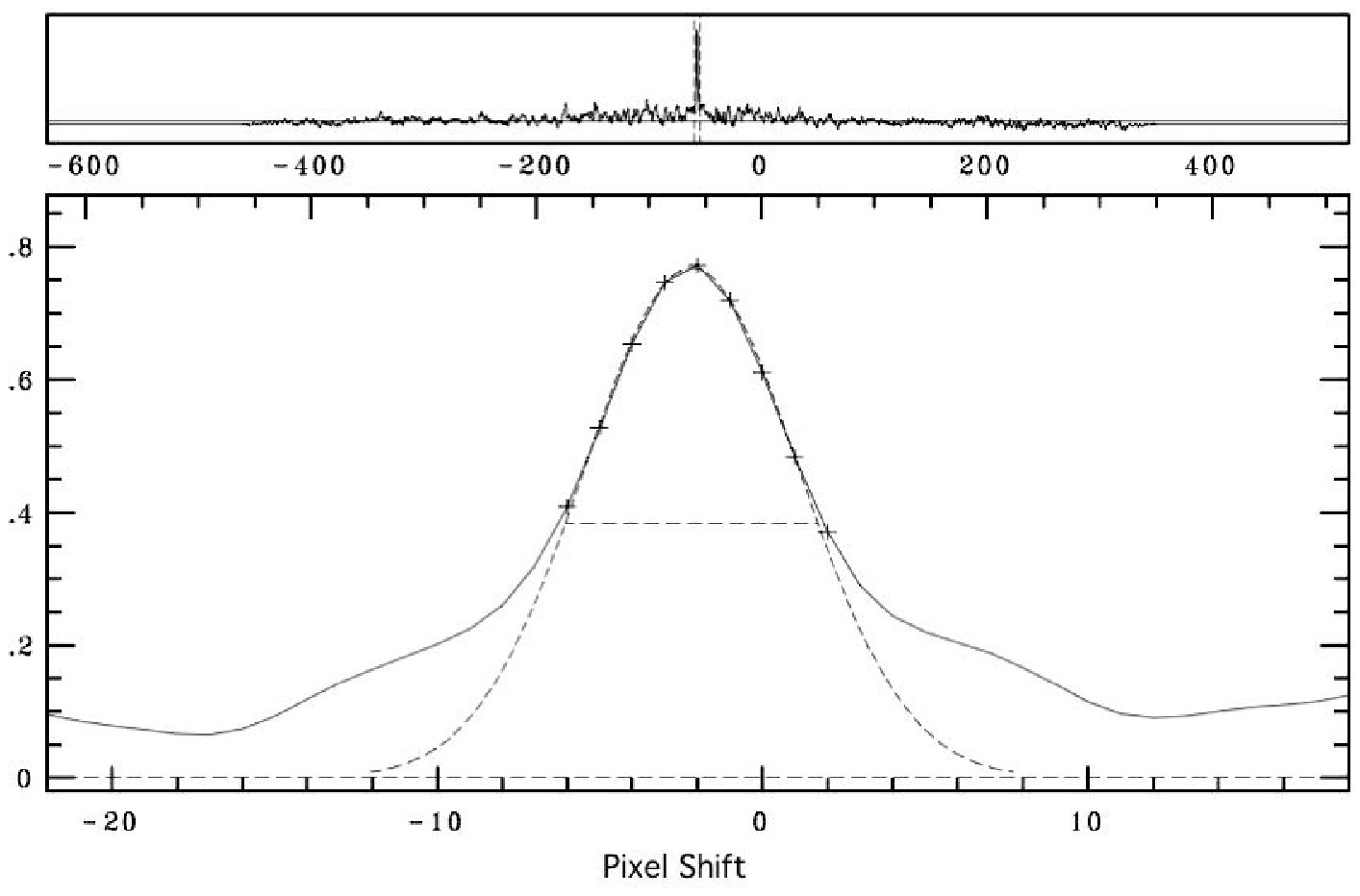}
\epsscale{0.35}
\plotone{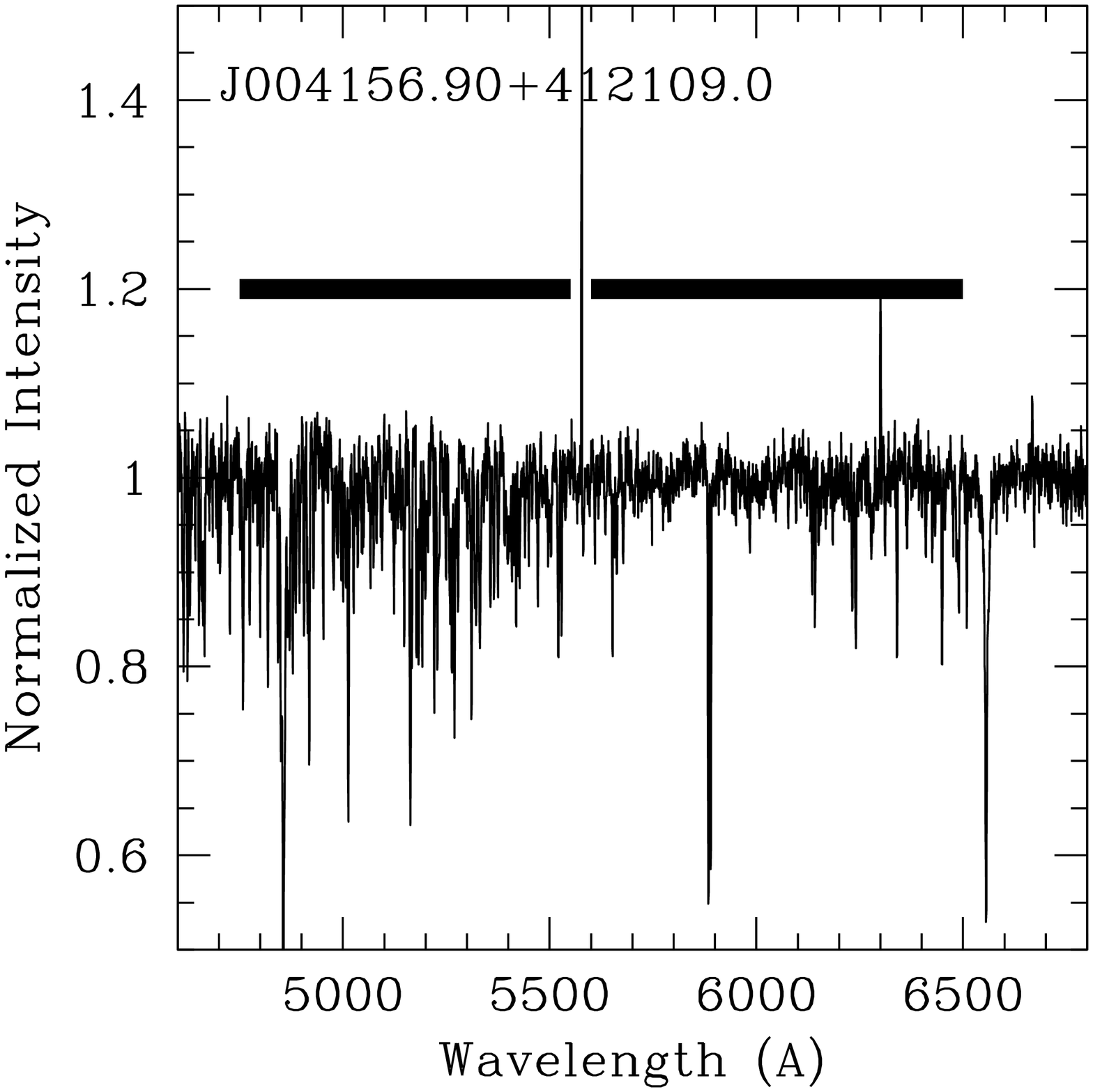}
\epsscale{0.55}
\plotone{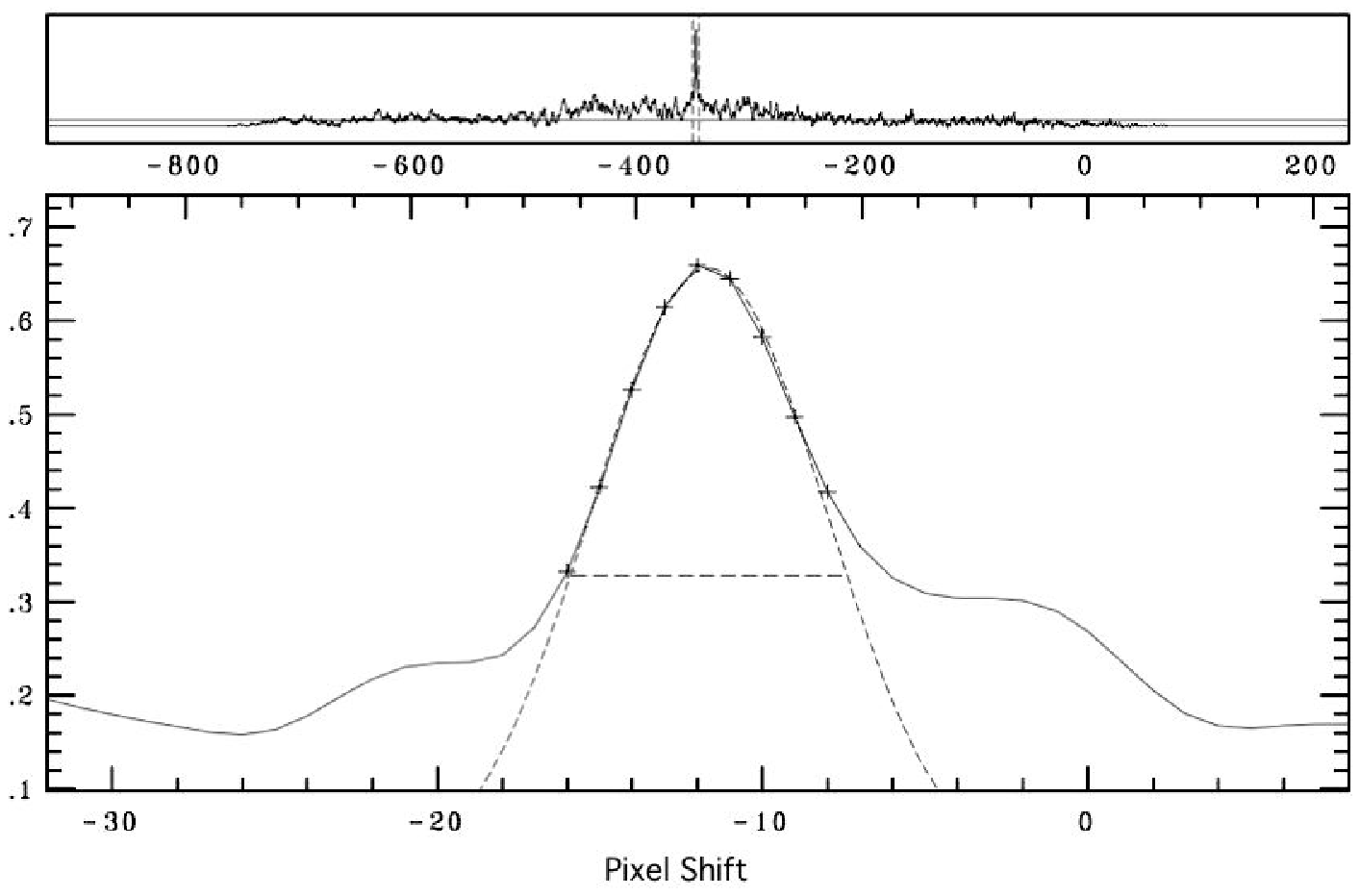}
\caption{\label{fig:spectra} Examples of spectra and cross-correlations.  We show sample spectra and the corresponding
cross-correlation functions and fits. The thick bar shown on the spectra indicates the 4750-5550\AA, 5600-6500\AA\ regions
used for the cross correlation. The upper star, J00303.99+395939.5, proves to be a foreground dwarf (\S~\ref{Sec-how}), with an average radial velocity of
$-48$ km s$^{-1}$ compared to -534 km s$^{-1}$ expected for its position. The lower star, J004156.90+412109.0, proves to be an M31 supergiant (\S~\ref{Sec-how}),
with an average radial velocity of $-335$ km s$^{-1}$ compared to $-303$ km s$^{-1}$ expected for its position.  Additional spectroscopy,
discussed in \S~\ref{Sec-OI}, confirms that the star has very strong OI $\lambda 7774$, characteristic of a yellow supergiant.
}
\end{figure}

We show in Figure~\ref{fig:spectra} examples of two of our spectra, and their resulting cross-correlation functions and corresponding
fits.  These were selected to have $V$ magnitudes typical of the median value in our entire sample, 17.1.

We list in Table~\ref{tab:all} the observed radial velocity of each of our 2889 objects, some of which had multiple observations.
What are the corresponding errors?  Each observed radial velocity $vel_{\rm obs}$  is the average of cross-correlating the star's spectrum against that
of the five radial velocity standard spectra.  The standard deviation of the mean of these is always very small, of order a few tenths
of a km s$^{-1}$.  However, the signal-to-noise in the standard templates is extremely high, and the standards are of similar type, so the differences in the velocities obtained from cross-correlating the same spectrum against these different standard spectra will certainly
underestimate the true uncertanty.  Tonry \& Davis (1979) instead introduce the $r$ parameter, which is the ratio of the peak of the correlation function to its noise.  
We can estimate the relationship between $r$ and the error by using measurements obtained for stars observed multiple times.
The 128 objects in the Brt5-1 field were observed on two separate nights, and in addition, there
were 28 objects 
in common between the ``bright" and ``faint" fields, plus 59 objects observed in multiple ``faint" fields.
We find that in general our velocity error (in km s$^{-1}$) is given by
 $${\rm Vel}_{\rm err} = 2.3 +11.5/(1+r), $$
 where the functional form reflects both errors in the standard star velocities and in uncertainties due to the cross-correlation
 (see Tonry \& Davis 1979).
 The typical (median) $r$ value for our data is 33, corresponding to a 2.6 km s$^{-1}$ error.
 
\begin{figure}
\epsscale{0.9}
\plotone{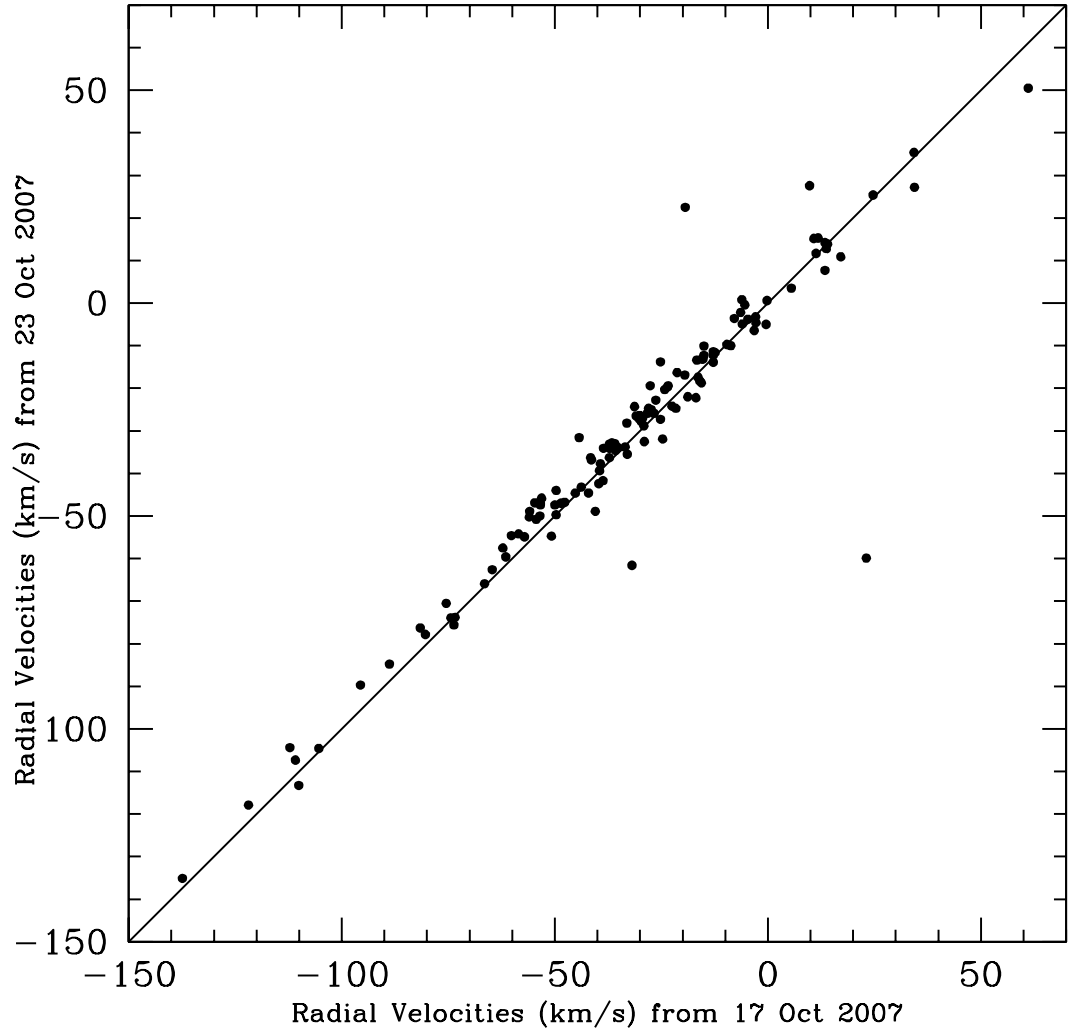}
\caption{\label{b5compare} Radial velocity constancy.  We show here a comparison of the
observed radial velocities for the two observations of the Brt5-1 field.  The line shows the 1:1
relation.}
\end{figure}

As a further check,  in Figure~\ref{b5compare} we compare the velocities obtained from the two
observations of the Brt5-1 field.  
We find that there is an excellent match between the observations over all the velocities we've sampled.  As mentioned above,
this was not the case when we originally used the low-dispersion templates, where we saw a velocity-dependent problem.
 For all future purposes in the paper, the velocities
 produced during multiple observations were averaged to yield one Vel$_{\rm obs}$ for each program object.
 It is clear that a few stars have variable velocities based on this comparison; we expect these stars to be binaries.
 If two observations of the same star differed by 10 km s$^{-1}$ or more we note this fact in Table~\ref{tab:all}.

\begin{figure}
\plotone{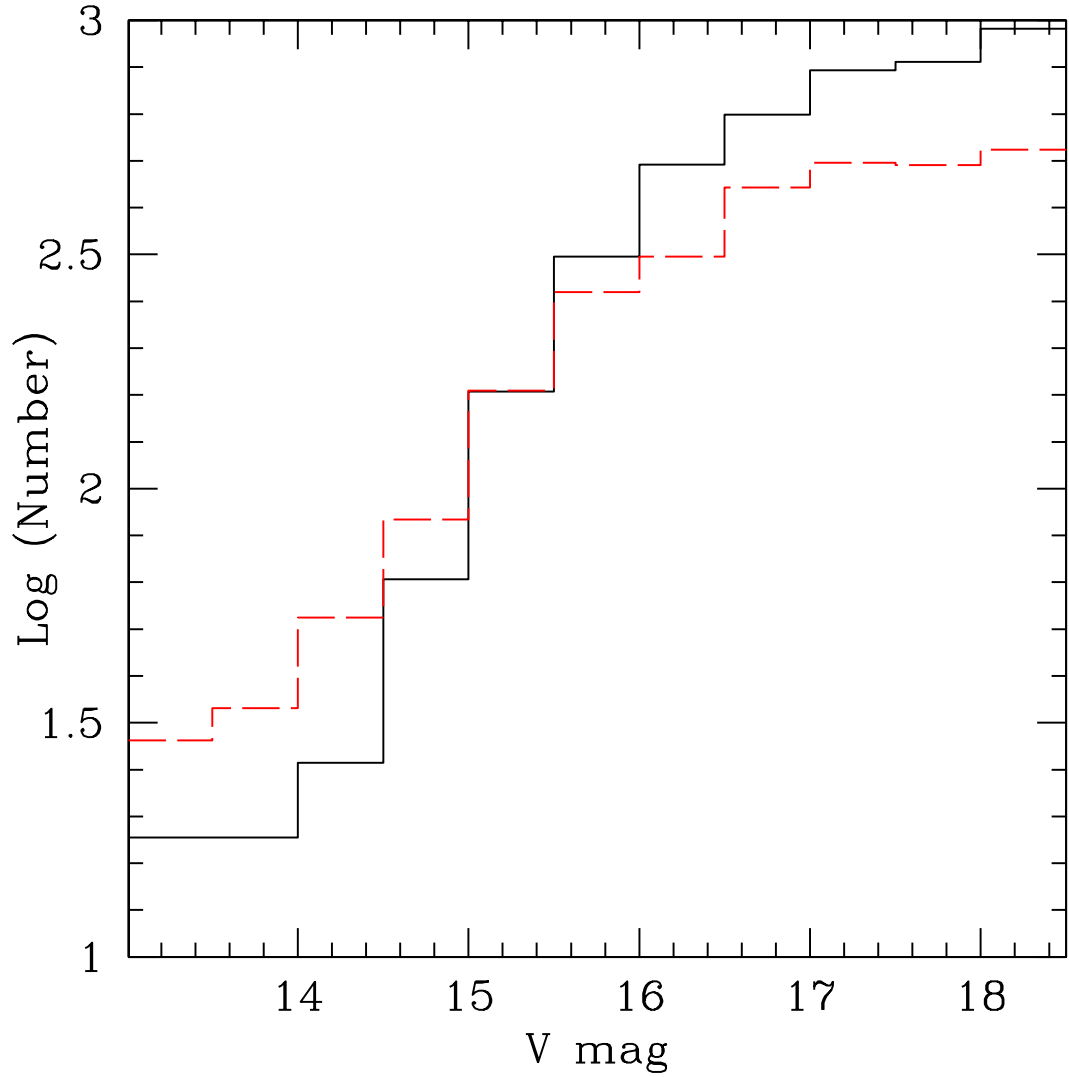}
\caption{\label{fig:magcomp} Distribution in magnitude.  We compare the magnitudes of
the subsample of stars for which we obtained radial velocities (red, dashed histogram) with that
of the parent population (black, solid histogram).
}
\end{figure}

Since we will be using the supergiants we identify to test if the models predict the same relative
number with luminosity, we should understand what magnitude biases exist, if any, in the 
subsample of targets for which we obtained radial velocities compared to the parent sample.   
In Figure~\ref{fig:magcomp} we show the distribution of magnitudes in the complete sample of 
4282 objects which met our original criteria (magnitude, color, and location) in black.  The red (dashed)
histogram shows the sample of 2989 objects for which we successfully obtained radial velocities.
We show this on a log plot so that a linear difference corresponds to a percentage; i.e., 
a difference of 0.3~dex is a factor of 2 regardless of the absolute numbers.  We find very good agreement,
with only a slight tendency towards having obtained radial velocities for a larger percentage of brighter stars
than fainter ones.  This will  introduce a slight bias in our final results in that
higher luminosity stars should be proportionally over-represented in our sample. 

\subsection{Identifying the Supergiants}
\label{Sec-how}

Now armed with the observed radial velocities of our 2899 program objects, two main steps are necessary to identify
 the M31 yellow supergiants.  First, foreground dwarf contaminants must be eliminated by 
determining which objects' velocities match that expected from M31's rotation curve and, second, once a list of M31 members has
 been obtained, any non-stellar objects (i.e., small clusters) must be removed.

\begin{figure}[ht]
\epsscale{0.6}
\plotone{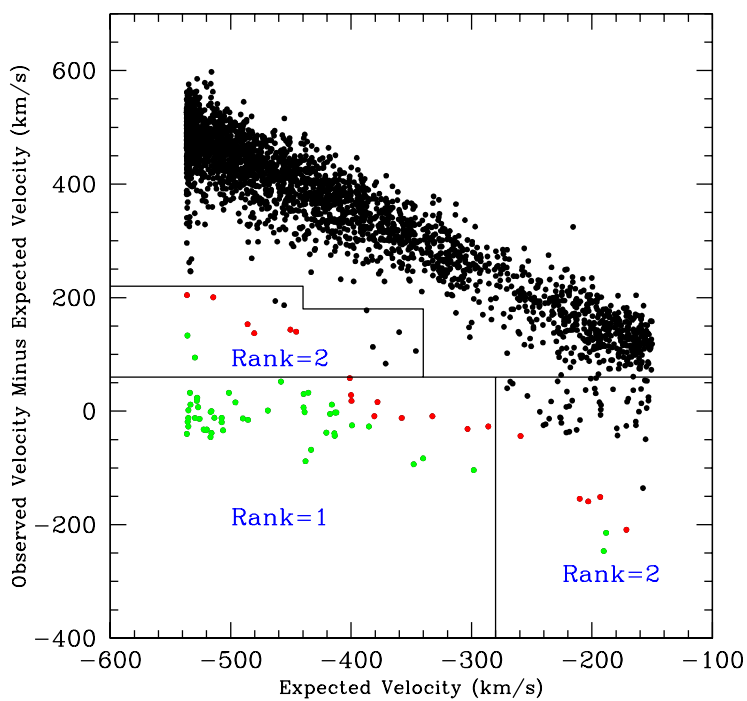}
\caption{\label{expectvdif}  Comparison between the observed and expected radial velocities.
Here we plot the difference between the observed and expected radial velocities vs.\ the expected
radial velocities.  Any M31 
members will lie near a difference of 0, while the dark band is composed of foreground
dwarfs. Those objects with V$_{\rm obs}$ $\leq$ -400 km s$^{-1}$ are marked in green and
 those with -400 km s$^{-1}$ $\leq$ V$_{\rm obs}$ $\leq$ -300 km s$^{-1}$ as red.The assigned ``ranks" (1= mostly certain
 supergiant; 2=probable supergiant) are shown.}
\end{figure}

In the notation of Rubin \& Ford (1970), the radial velocity $V_r$
of an object in a disk galaxy can be approximated by $V_r=V_0+V(R) \sin{\xi} \cos{\theta}$
 where $V_0$ is the systemic radial velocity,
 $\xi$ is the angle between the line-of-sight and the perpendicular to the plane of the galaxy, V(R) is the rotation
 velocition within the plane at a radial distance R, and $\cos{\theta} = X/R$ where X is the position along the major axis.
  This would be a linear relationship only if the rotation curve was absolutely flat i.e. V(R)= const.  However, as mentioned
 in \S~\ref{SS} this is a good approximation for M31.  We used the same fit, as described there, to compute what the 
radial velocity of each of our program objects \emph{should have been} (given its position on the sky)
were it an M31 member.  The difference between our
 V$_{\rm obs}$ and V$_{\rm expect}$ for all objects was then calculated and plotted against V$_{\rm expect}$, as shown in Figure~\ref{expectvdif}.

In Figure~\ref{expectvdif}, all the objects whose velocities correspond with M31's rotation curve should lie along the zero
 point of the Y axis.  The left hand side of the plot, where the expected velocities are highly negative, represents the SW portion
 of M31, rotating \emph{towards} the Milky Way.  Thus, the strong diagonal band represents the foreground dwarfs (objects
 with essentially zero radial velocity).  As can be seen, the M31 members can easily be distinguished from the foreground dwarfs
 on the left side of the plot and the distinction becomes increasingly hazy as we move along M31's semi-major axis.  We therefore
 distinguish between two ``ranks" of M31 members.  Rank 1 objects are those which we can say  are ``nearly certain" M31 members, 
whereas rank 2 objects we consider to be  ``probable"  M31 members.  All objects with an expected velocity $< -280$ and difference
 $< 60$ were labeled as rank 1, and those with expected velocity $> -280$ and difference $< 60$, expected velocity $< -340$ and 
difference $< 180$, or expected velocity $< -440$ and difference $< 220$ were labeled as rank 2.  These distinctions are displayed
 in Figure~\ref{expectvdif}.

As a result of this classification we were left with a list of 56 nearly certain  and 71 probable M31 members.   There still existed
 the possibility, however, that some of these M31 members were not stars, but small clusters.  This
was addressed by measuring the size of the objects as compared to nearby stars.  We did this using two methods, both utilizing the 
images of the LGGS. Initially the 127 potential supergiants were examined with the 
Source Extractor (Bertin \& Arnouts 1996), using the V-band images, and the results were then compared to an independently conducted
 manual check on the R-band.  The manual check was completed by measuring the FWHM of each potential supergiant and
 its neighbors in the
 IRAF package imexam.  The agreement between the manual and Source Extractor method was complete with each yielding seven previously
 known (Galleti et al.\  2007) clusters in our sample: Mag-237751, J004345.23+410608.5, J004446.42+412918.3, J004545.58+413942.4, 
J004356.46+412203.3, J004358.15+412438.8, and J004403.98+412618.7.   

Once the foreground contaminants and M31 clusters had been removed from our sample and our rank
 designations applied, we found that, out of our original sample of 2899 objects, we are left with 54 rank 1 and 66
 rank 2 yellow supergiants.  This corresponds to a foreground contamination between 96-98\%! 
The velocity information for these 120 objects, as well as the 
seven M31 clusters (listed at the end) is summarized in Table~\ref{tab:observed}.  

How clean is our remaining list of candidates?    As discussed in \S~\ref{Sec-wheat} we expect that our original 
sample to contain some small fraction of 
Milky Way halo stars in addition to the numerous foreground disk stars.   Although the radial velocities
are effective at eliminating the disk contaminants, they will be less effective at weeding out the halo stars.
Of the $\sim -300$ km s$^{-1}$ systemic velocity of M31, about two-thirds of  
it is the reflex motion of the sun: equation 4 of Couteau \& van den Bergh (1999) implies a reflex motion of $-178$ km s$^ 
{-1}$.   Stars from the halo are therefore likely to reflect this solar motion.
If the halo's velocity dispersion is 130 km s$^{-1}$ (Binney \& Merrifield 1998), then 3$\sigma$ velocities  
could extend all the way to -570 km s$^{-1}$. If there was a significant number of such stars in our sample, then  
some contamination could occur.

\begin{figure}[ht]
\epsscale{.5}
\plotone{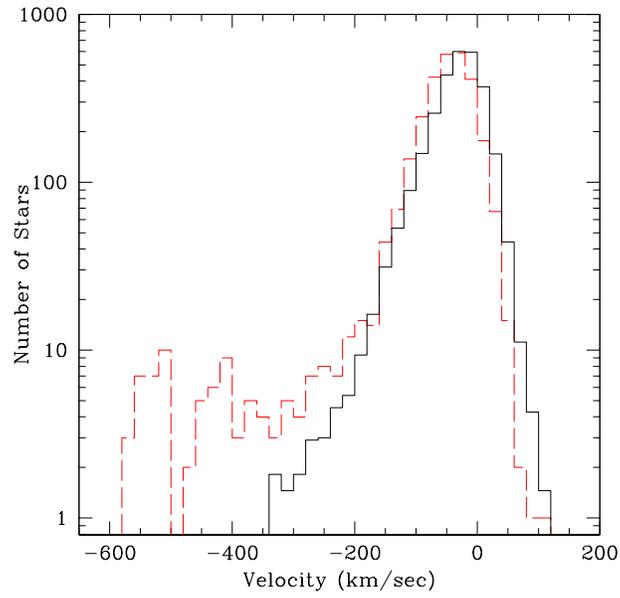}
\caption{\label{histo2} Histogram of radial velocities.   The observed radial velocities (red, dashed histogram) is compared
with what we expect for the foreground contamination (black, solid histogram) computed from the Bescancon model.
}
\end{figure}

 In Figure~\ref{histo2} we compare our observed velocities (dashed, red histogram) with that
expected for foreground stars according to the Besancon model (solid, black histogram). 
We see that virtually all of the stars with observed velocities greater than $-175$ km s$^{-1}$ are  
foreground objects, as expected.  At $-200$ km s$^{-1}$ about half of the objects should be foreground,  
and half M31.  More negative than this, the M31 population dominates, with  
increasingly little foreground contamination. The detailed output of the Besancon model shows 7 foreground stars with radial  
velocities $\leq -300$ km s$^{-1}$, and 1 foreground star with a radial velocity $\leq -400$ km $^{-1}$.

We can also estimate this contamination independently using the Bahcall \& Soneira (1980) model which 
predicts about 40 halo dwarfs per square degree within the magnitude and color range we use in the direction of M31.
As for halo {\it giants},  the Bahcall \& Soneira (1980) model overestimates their  number (Morrison 1993), but Heather  
Morrision (private communication) has been kind enough to calculate that there should be about four such objects in
a square degree seen towards M31 in this same magnitude and color range.  The effective area in our survey is about 1 deg$^2$, 
and  in it we observed 68\% of the stars in our original sample, so we might realistically expect about 30  halo stars in our spectroscopic sample.
Assuming the radial velocity distribution is Gaussian, this leads to the identical results as above: there should be 5 halo stars 
with velocities more negative than $-300$ km s$^{-1}$ (0.94$\sigma$) and 1 below $-400$ km s$^{-1}$ (1.8$\sigma$). 

So, simply noting that the rank 1 stars all have have velocities more negative than $-280$ km s$^{-1}$, we expect {\it at most} 8 stars in
this sample of 54, or 15\%.  However, this is likely an overestimate of the contamination, as it  ignores the additional information gained by the star's 
position and therefore its \emph{expected}  radial velocity, V$_{\rm expect}$, were it an M31 member.  To be a rank 1 candidate, a star has to be in
the right   location within the M31 field such that its V$_{\rm obs}$ corresponds to its V$_{\rm expect}$ 
(which, although still possible, is improbable).  We therefore refer to the rank 1 candidates as ``nearly certain".   The situation for the rank 2
candidates is harder to evaluate, as a few of these have radial velocities as positive as $-100$ km s$^{-1}$, but their location in the
Figure~\ref{expectvdif} leads us to consider them ``probable" yellow supergiants.  Contamination by foreground stars of half of this
sample would not surprise us, however.

Of our original 2899 program objects, four had previously appeared in the literature as spectroscopically confirmed members
 of M31.  J003745.26+395823.6 appeared in Humphreys (1979) as IV-A207 and was classified as F5 Ia, J004101.24+410434.6 
can be found by the name OB69-46 in Massey (1998a) as a red supergiant (based on earlier photometry), 
and J004129.31+405102.9 is listed as OB22A in Humphreys et al.\ 
(1990) and classified as a F8-G0 Ia.  We categorized all three as rank 1 yellow supergiants.  Additionally, Humphreys
 et al (1988) lists J004101.55+403432.3 (as III-R61) as a M31 RSG (K5 I) candidate. However, we are forced to conclude
 that this star is actually a foreground dwarf, given its $-41$ km s$^{-1}$ radial velocity (at its position in M31, the expected
 radial velocity is $-464$ km s$^{-1}$.  We have double checked our previous cross-identification of this star, and it matches
 the one shown on the Humphreys et al.\ (1988) finding chart; the colors $B-V=0.69$ and $U-B=0.02$ do not correspond to mid K-type,
 and the spectrum is clearly of earlier type, with the Balmer lines prominent, along with strong lines of Mg I and Fe I.

\subsection{Physical Properties: Transforming the Photometry}

In order to use our list of M31 yellow supergiants to test the current stellar evolutionary tracks,
 it is necessary to determine their effective temperatures and bolometric luminosities.  For this, we will transform
 each star's photometry, as described below.   

The $B$ and $V$ photometry for all but one of our stars came from
 the LGGS; that of the single ``bright" star, Mag-253496, is taken from Magnier et al.\ (1992) and adjusted by the 
 small correction found by Massey et al.\ (2006).   We apply a constant reddening correction $E(B-V)=0.13$ based on the
 median value found for early-type stars in M31 by Massey
 et al.\ (2007a), and is in accord with the color excess derived from spectral types of a handful of O-type stars (Massey et al.\ 1986).
  The reddening of individual stars can readily differ from this (by several tenths) but in this part of the CMD 
 reddening-free indices such as $Q$ are degenerate with $T_{\rm eff}$, as one may notice in Figure~\ref{fig:2color}, the reddening vector is nearly parallel to the supergiant
 sequence.
 
The problem now becomes how to best translate these dereddened colors into effective temperatures.  
Flower (1996) and  Kovtyukh (2007), both present empirical effective temperature scales that include F- and G-type
supergiants.   The
Flower (1996) data is drawn from the literature, while Kovtyukh (2007) performs his own analysis on spectra of Galactic
supergiants.  The problem is that both of these studies are based upon samples of Galactic supergiants,  while we expect the
metallicity of our M31 stars to be about $2\times$ solar based upon HII region abundances (i.e.,  Zaritsky et al.\ 1994).
We therefore have decided to instead use the Kurucz (1992) ``Atlas9'' model atmospheres to provide the transformations.

\begin{figure}[ht]
\epsscale{0.5}
\plotone{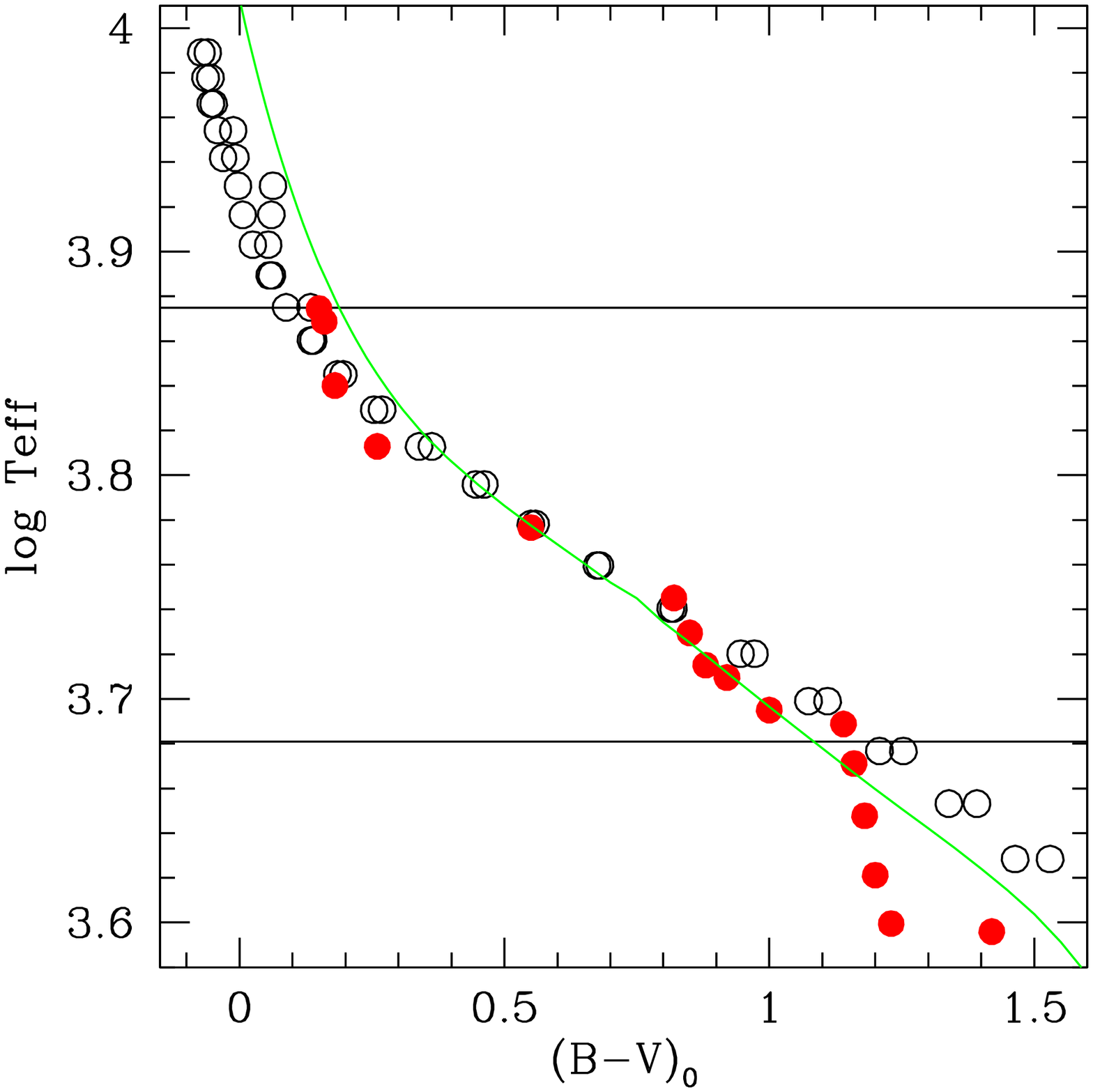}
\caption{\label{Atlas} Comparison of effective temperature scales.  The Atlas9 model predictions are shown by open (black) circles for 
surface gravities appropriate to supergiants.  The filled (red) circles are from the Kovtyukh (2007) study of F- and G-type supergiants.
The solid (green) curve is the Flower (1996) relation for supergiants.   the two horizontal lines correspond to effective temperatures of
7500~K and 4800~K, the region we are considering the realm of the yellow supergiants here.}
\end{figure}

In Figure~\ref{Atlas} we compare the Altas9 models for solar metallicity with the two empirical calibrations\footnote{We have used the intrinsic colors of FitzGerald (1970) as a function of spectral types
in order to assign $(B-V)_0$ values to the spectral types given by Kovtyukh (2007).}. For this comparison, we have used the lowest
two surface gravities for each temperature model computed by Kurucz (1992), i.e., $\log g=0.0$ and 0.5 for $T_{\rm eff}\leq 6000$~K, $\log g=0.5$ and 1.0 for 6250~K$\leq T_{\rm eff} <$ 7500~K, and $\log g=1.0$ and 1.5 for
7500~K$\leq T_{\rm eff} \leq$ 8500~K. For the purposes of comparing our data to the stellar evolutionary models, we will restrict ourselves only to our defined yellow supergiant effective temperature range: 7500~K to 4800~K ($\log T_{\rm eff}=3.875$ and 3.681). This region is indicated by the two horizontal lines in Figure~\ref{Atlas}.  We
can see that over this temperature range there is substantial agreement between the models and the empirical calibrations.

As mentioned above, one advantage of using the Atlas9 models is that we can fine-tune the transformations to an appropriate metallicity, although we will find below that this correction is
very slight. In keeping with Zaritsky
 et al.\ (1994) we assume an M31 metallicity of $2\times$ solar (see further discussion in Crockett et al.\ (2006) and Massey et al.\ 2009).
 Using the Atlas9 model with the most similar metallicity ($1.6\times$ solar) we compute a relationship between $(B-V)_0$ and
 $\log T_{\rm eff}$ as follows:
 $$\log T_{\rm eff}=3.913-0.3512(B-V)_0+0.2692(B-V)_0^2-0.1108(B-V)_0^3$$
 
 To keep the relationship accurate over the range of temperatures in which we are the most interested, we restricted the fit to models that
 just bracketed the temperature range above (i.e., Kurucz models with 4750 K$\leq T_{\rm eff} \leq$ 8000 K), and therefore this relationship
 is {\it only} applicable for $0.03\leq (B-V)_0 \leq 1.26$.  Since we adopted an average value of $E(B-V)=0.13$, this then corresponds to
 $0.16\leq B-V \leq 1.39$, a good match to the $0.0 \leq B-V \leq 1.4$ of our sample.   Only six stars of our 120 yellow supergiants have colors bluer than
 $B-V=0.16$.   Although we will not include these in our comparison to the evolutionary models in the next section, we nevertheless would like to include these
 in the H-R diagram, and so we will adopt the following approximate transformation for these stars: $\log T_{\rm eff}=3.934-0.549(B-V)_0$.
 
 The bolometric corrections are relatively modest for yellow supergiants (a few tenths of a magnitude), and we derive the following relationship using the 
 Atlas9 models; the results are valid btween 4750~K to 9500~K, and, hence, applicable to our complete sample:

 $${\rm BC}= -251.54 +130.763\log T_{\rm eff} - 16.9934(\log T_{\rm eff})^2$$

 How much difference does adopting $1.6\times$ solar metallicity make?   The difference is slight: for a star with
 $(B-V)_0=0.6$ (roughly corresponding to 6000~K) we would derive $\log T_{\rm eff}=3.775$ using the $1.6\times$ Atlas9 relationship given above,
 while we would derive $\log T_{\rm eff}=3.769$ from an analogous relationship derived from the $1.0\times$ models.   The difference, 0.006~dex, is negligible.
 
 We give the derived physical properties in Table~\ref{tab:derived} for the 120 probable supergiants sorted by bolometric luminosities. A distance modulus of 24.40 (0.76~Mpc) was adopted,
 following the discussion in van den Bergh (2000). 

\subsection{Membership Re-examined}
\label{Sec-OI}

One of the critical and interesting properties we are attempting to determine is the upper luminosity limit for yellow supergiants, and we were struck by the fact that the
ten most luminous stars in Table~\ref{tab:derived} were all of rank  2.  We argue above that we expect some
(minimal) contamination by foreground (halo) giants in our sample, as these stars would have radial velocities characteristic of the reflex
motion of the sun.  In particular, the three most luminous possible supergiants stand out as extraordinarily bright.  None of the three have
extreme radial velocities (which is in part why they are rank ``2"): Mag-253496 has a Vel$_{\rm obs}$ of -227 km s$^{-1}$, J004251.90+413745.9 has
a Vel$_{\rm obs}$ of -210 km s$^{-1}$, and J004618.59+414410.9 has a Vel$_{\rm obs}$ of -150 km s$^{1}$, as can be seen in
Table~\ref{tab:observed}.   We were therefore very
keen to confirm or refute their membership in M31. The wavelength range of our own spectra had been optimized for radial velocities and where
Hectospec has good throughput,
and did not include the various good luminosity indicators in the far blue (see below) nor the O I$\lambda 7774$ triplet
in the far red.

The Kitt Peak Director was sympathetic to our plight and arranged follow-up spectra to be obtained for these three stars plus comparison
spectral standards.   The data were obtained as part of engineering time on 
the 3.5-m WIYN telescope with the Hydra fiber spectrometer in the far red, and on 
the Kitt Peak 4-m Mayall telescope with the RC spectrograph (in the blue). 
These spectra convinced us that all three of these are actually halo stars, and not M31 supergiants, as we argue below; they also eliminated another rank 2 star from membership, and confirmed
membership of other (both rank 1 and 2) stars.

The WIYN Hydra spectra were obtained on 11 August 2008, and consisted of an 1800 sec exposure and a 1300 sec exposure; the latter was ended
due to clouds.  The setup included Mag-253496 and J004251.90+413745.9 and a number of other stars, but not J004618.59+414410.9.   The (2.8 pixel) resolution was  4.0~\AA.
The wavelength range included the O I$\lambda 7774$ triplet, known to have a strong luminosity effect in F-type supergiants
(Osmer 1972) due to non-LTE effects, exacerbated by sphericity and the large mass outflows found
in supergiants (Przybilla et al.\ 2000).  We include in Table~\ref{tab:derived} what we find for this line.  The two most luminous supergiant
candidates have little or no O I $\lambda 7774$, arguing they cannot be supergiants, and we identify another non-supergiant
among the rank 2 objects we observed.  Strong O I $\lambda 7774$ is found for several of the other rank 1 and rank 2 candidates.

The 4-m RC Spectrograph spectra were obtained on 4 September 2008, and consisted of a 3x900 s exposure of Mag-253496, and 3x500 s 
exposures of J004251.90+413745.9 and J004618.59+414410.9.  The wavelength region included 3880-4600~\AA\ at a (2.2 pixel)
resolution of 1.6~\AA.  Several spectral standards were also observed to provide guidance in interpreting the data; these were supplemented
by similar data obtained at a later data with the Kitt Peak 2.1-m Goldcam spectrometer at a similar dispersion.  These data show that
Mag-253496 is roughly of spectral type G8, based upon the  Fe I $\lambda 4143$/ H$\delta$ and Fe I $\lambda 4045$/H$\delta$
ratios (Keenan \& McNeil 1976).  J004251.90+413745.9 and J004618.59+414410.9 are of earlier type, between F5 and G2, based upon
the strength of the G-band compared to H$\gamma$, and we adopt an F8 type.

For the G8 star, Mag-253496, we find that the ratio of Sr II $\lambda 4077$ to the Fe II/III blend at $\lambda 4063$ is about 1,
typical of an G8 III; in a supergiant, this ratio would be considerably larger (e.g., Keenan \& McNeil  1976). Similarly, the ratio of the Fe, SrII
blend at $\lambda 4216$ to the Ca II $\lambda 4226$ is quite small, consistent with that of a dwarf or a giant, but not a supergiant.
  Combined with the lack of O I $\lambda 7774$, we therefore
conclude this star is a Milky Way halo giant, and not an M31 supergiant.  For the F8 stars, we find that the strengths of TiII $\lambda 4444$
and Mg II $\lambda 4481$ to be roughly equal, consistent with a giant, but inconsistent with a supergiant.  Add to that the lack of O I $\lambda 7774$
in J004618.59+414410.9, and we conclude these are also halo giants.  

We also include in Table~\ref{tab:derived} the results of 6 other stars for which previous spectroscopy by one of us (N. C.) had identified
strong O I $\lambda 7774$.  Five of these are rank 1, and one is rank 2.  (An additional two stars, one of rank 1 and one of rank 2, were
in common with the WIYN spectroscopy, and agreed with our assessment.)   These were all obtained with Hectospec using a lower resolution (5\AA)
grating as part of a different project (Caldwell et al.\ 2009).  
In the following, we consider all stars found to have strong O I $\lambda 7774$ as certain supergiants, even if they were rank 2 based on their
radial velocities, while we have removed the four stars shown spectroscopically not to be supergiants.

\begin{figure}
\epsscale{0.8}
\plotone{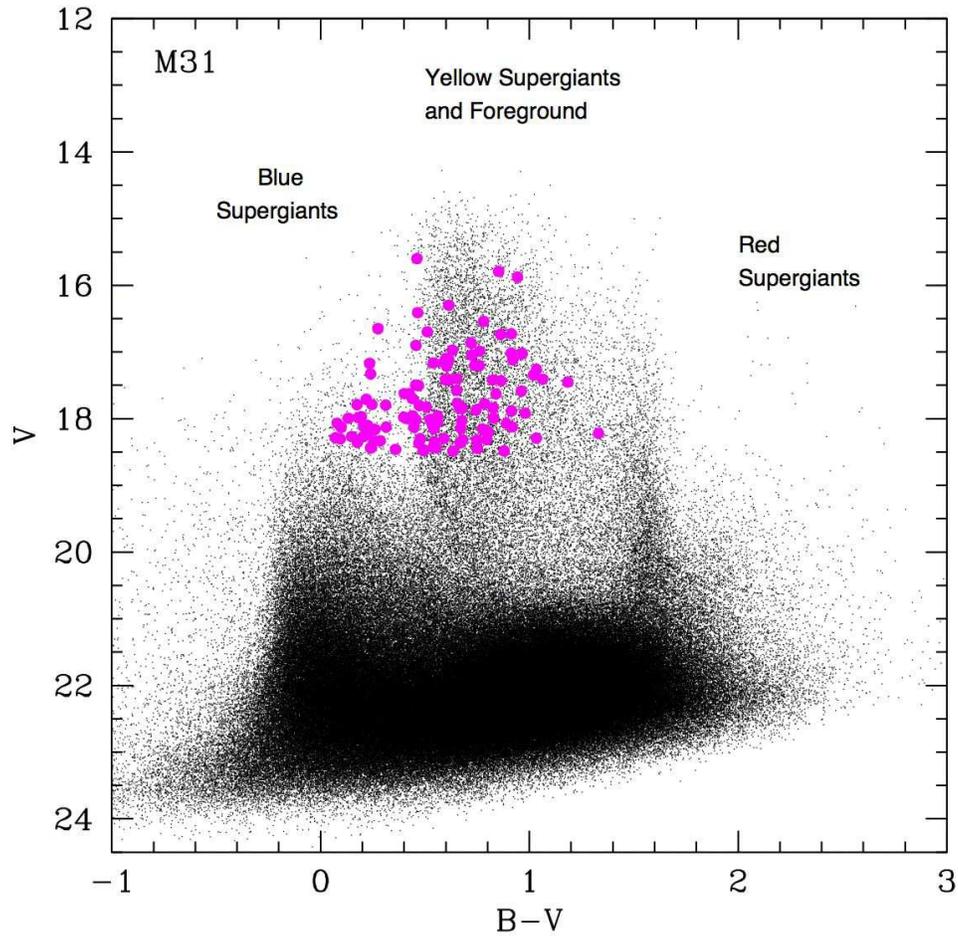}
\caption{\label{fig:CMD2} Color-magnitude diagram revisited.  Same as  Figure~\ref{fig:CMD}, but the addition of the newly confirmed yellow
supergiants marked as colored filled circles.
}
\end{figure}

It is of interest to see where our supergiants form in the various diagnostic diagrams we employed earlier.   First, in Fig.~\ref{fig:CMD2} we
superimpose the actual yellow supergiants on the CMD of M31.   Note that indeed our bluest ``yellow" supergiants extend into the blue supergiant
region, as expected, given our relaxation of our original color selection to include stars as blue
as $B-V=0$ (\S~\ref{SS}).

\begin{figure}
\epsscale{0.8}
\plotone{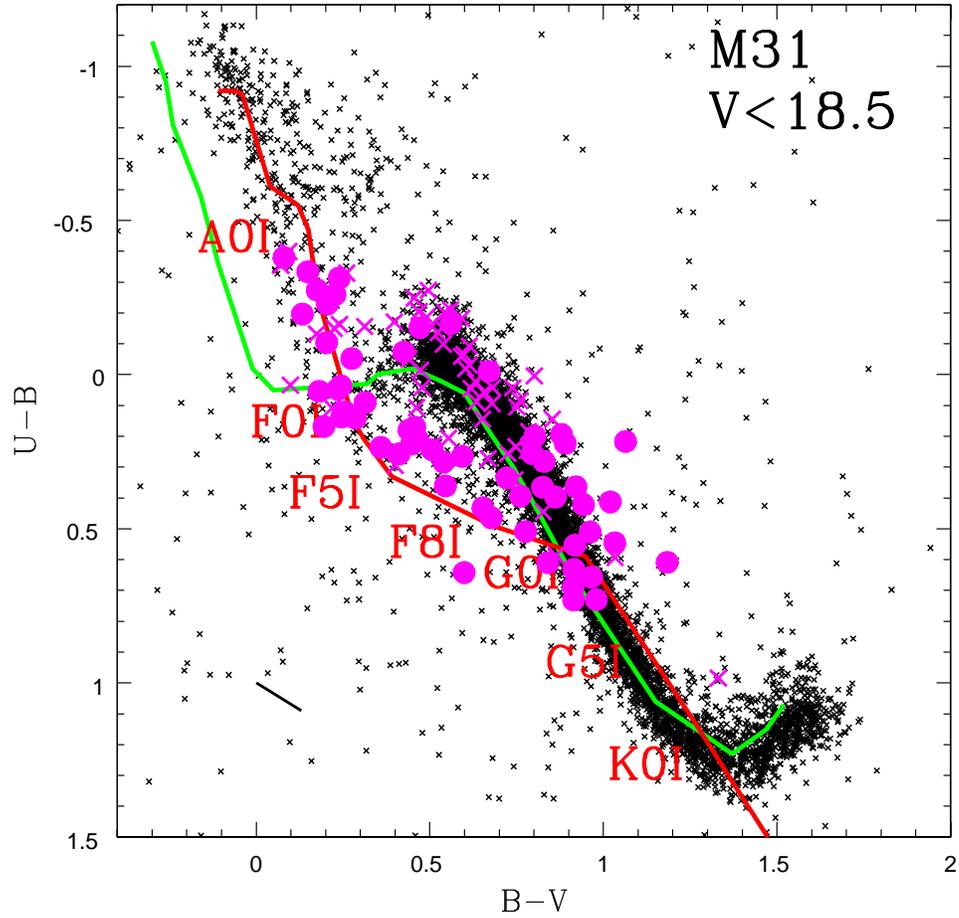}
\caption{\label{fig:2color3} Two-color diagram revisited.  Same as Figure~\ref{fig:2color} {\it left}, but with the yellow supergiants from this paper
now marked by filled circles (if rank 1 or spectroscopically confirmed rank 2) or as x's (if unconfirmed rank 2).  }
\end{figure}

We next show in Figure~\ref{fig:2color3} the locations of our yellow supergiants in the two-color diagram.   We see that, had we relied upon
a two-color plot to eliminate yellow supergiant candidates, we would have missed a number whose $U-B$ colors are
more negative than would be expected from the nominal supergiant sequence (shown in red).   Of course, it could be that some of
the stars  with $B-V>0.4$ but $U-B<0.1$ will turn out to be foreground upon additional spectroscopy as none of the four rank 1 stars
in this region yet have confirming O I $\lambda 7774$ spectroscopy.    We have examined the spatial location of these four stars, however,
and find that they are both in the spiral arms and near to other rank 1 yellow supergiants.  we, therefore, feel that it is more likely this is a combination of 
photometric errors, slightly variable reddening, and uncertainties in the two-color relation for yellow supergiants.

\begin{figure}
\epsscale{0.8}
\plotone{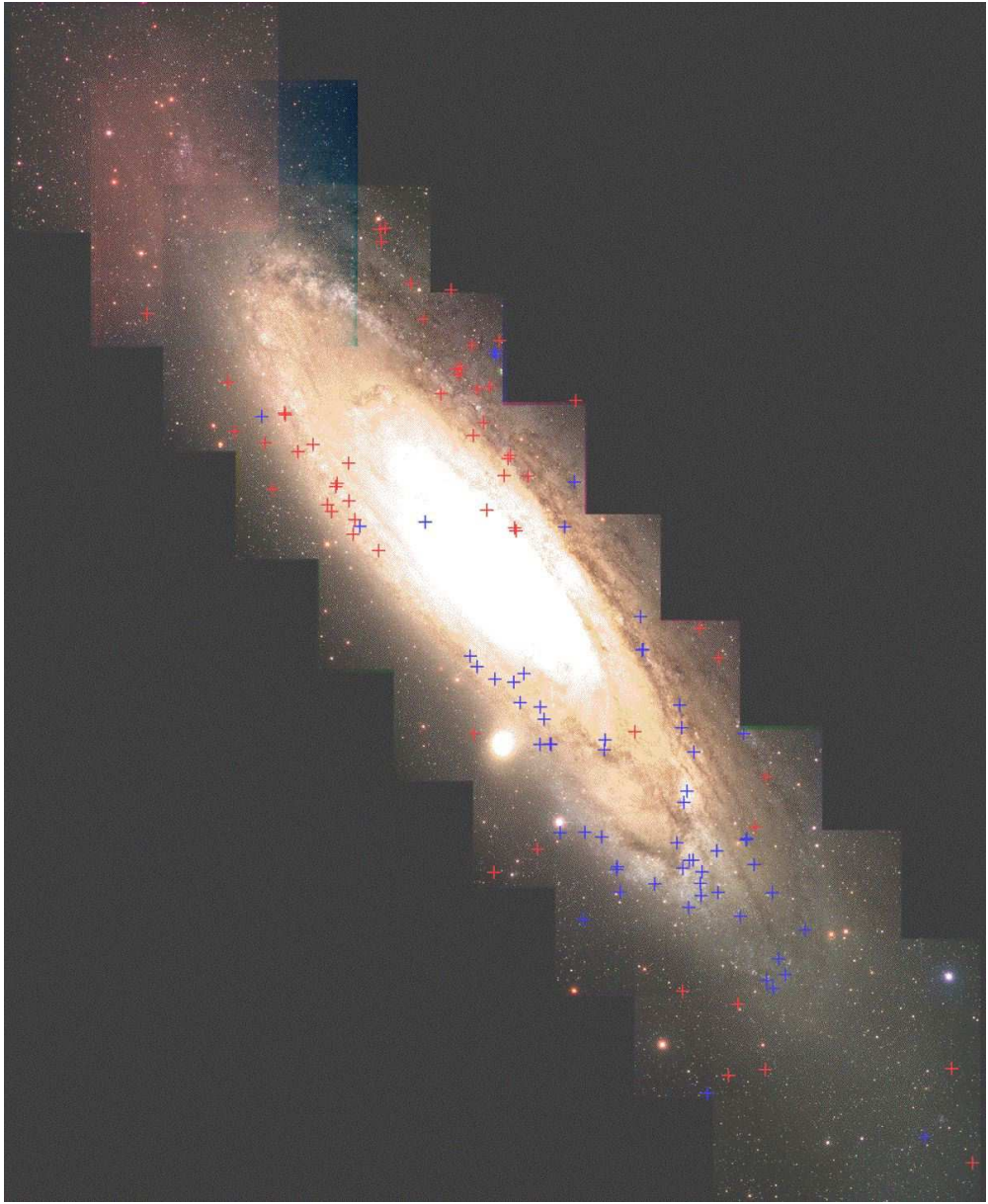}
\caption{\label{fig:winners} The distribution of our yellow supergiants across the face of M31.  The blue symbols represent the
rank 1 (certain) yellow supergiant candidates, while red represents the rank 2 (less certain) candidates. Compare with
Figure~\ref{fig:sample}.}
\end{figure}

 Finally, we show in Figure~\ref{fig:winners} the spatial distribution of our supergiant candidates in M31, where blue symbols indicate the certain ones
(either rank 1 or spectroscopically confirmed), and the red symbols the rank 2 candidates. For the most part, the yellow supergiants are found along the CO ring where star formation
is most active, as expected.

\section{\label{HR}The H-R Diagram}

\begin{figure}
\epsscale{.8}
\plotone{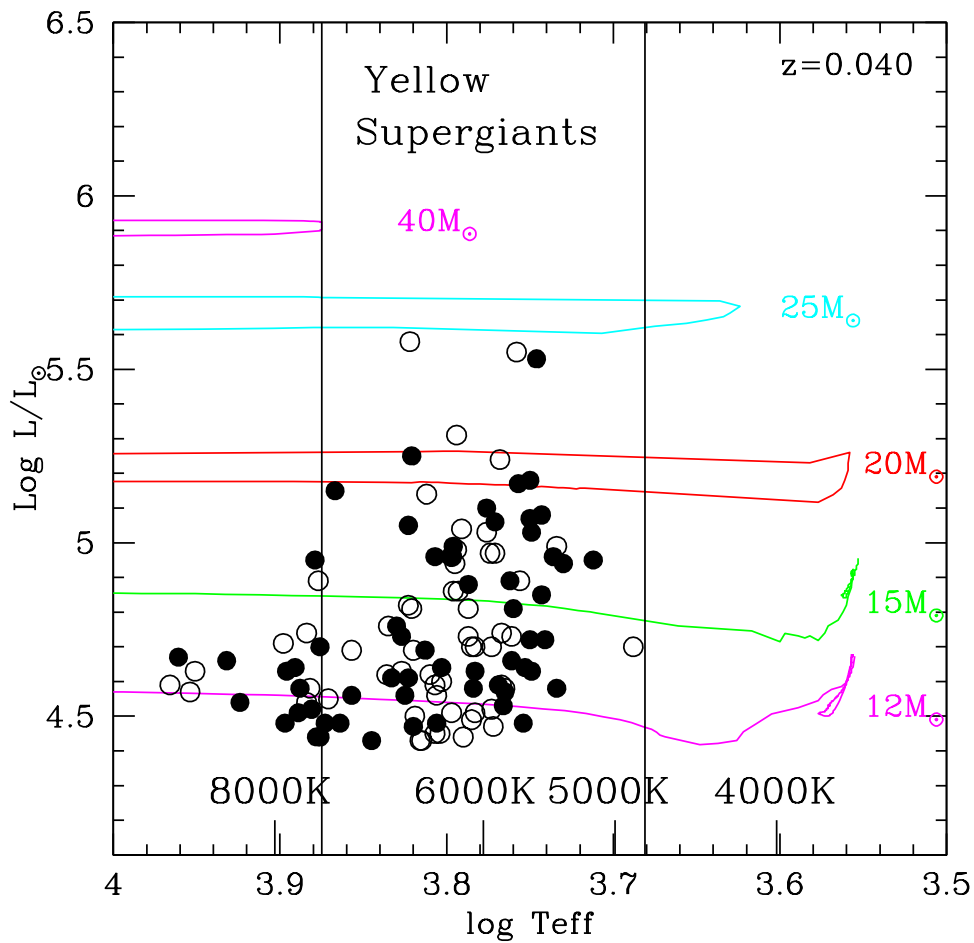}
\caption{\label{fig:HR} H-R diagram.  For simplicity, we show
only the latest ($z=0.040$) Geneva tracks with an initial rotation of 300 km s$^{-1}$; the location of the older
tracks can be seen in Figure~\ref{fig:models}. The solid points are
our certain yellow supergiants (either rank 1 or spectroscopically confirmed rank 2) while
the open points are the as-yet unconfirmed rank 2 yellow supergiant candidates. }
\end{figure}

In Figure~\ref{fig:HR} we show the location of our 116 yellow supergiant candidates in the H-R diagram, along with the
 $z=0.040$ evolutionary tracks.  For simplicity, we have shown only the newer models which 
include an initial rotation of 300 km$^{-1}$ (``S3" in Table~\ref{tab:ages}) as we view these as the most physically
realistic.  Several of these ``S3" tracks were computed specifically for this project, and that of Massey et al.\ (2009).   

First, we find that the tracks do a good job of predicting the locations of yellow supergiants in the HRD.  The most luminous yellow supergiants
in our sample have $\log L/L_\odot\sim 5.6$.   We do {\it not} find yellow supergiants with luminosities of (say) $\log L/L_\odot \sim 6$,
and this is in accord with what the evolutionary tracks predict.  Note that with the older tracks (dotted tracks in Figure~\ref{fig:models}) we
might have expected to see some higher mass, warmer yellow supergiants. Indeed, the number of high luminosity yellow supergiants
should have been similar to that seen  for lower luminosity 12-20$M_\odot$ yellow supergiants, as the older 60$M_\odot$ track extended into
the yellow supergiant realm, and the duration of the yellow supergiant phase was 5,400 years (comparable to that of the lower masses as seen in Table~\ref{tab:ages}).

This agreement with the new tracks is similar to what Massey et al.\ (2009) found for the coolest supergiants ($T_{\rm eff} \le 3800$ K) in terms of the
excellent agreement between the location of the tracks (and in particular the upper luminosities) and the observed locations of the
stars.  M31's RSGs have a maximum luminosity of $\log L/L_\odot \sim 5.4$, a little bit lower than the most luminous yellow supergiants,
as might be expected from the evolutionary tracks shown in Fig.~\ref{fig:HR}, as the 25$M_\odot$ track does not extend to such cool
effective temperatures.

We now tackle the test we described in \S~\ref{Sec-magglass}, namely to see whether the relative number of yellow supergiants increases
as we go to higher luminosities as the lifetimes in Table~\ref{tab:ages} suggest to us.  A visual inspection of Figure~\ref{fig:HR} says that
the answer is clearly no: the number of yellow supergiants decrease.  But, let us attempt a more quantitative assessment.

We list in Table~\ref{tab:numbers} the number of yellow supergiants we find in each mass bin, both for the
entire sample (rank 1 and rank 2) and for just the ones we are most certain are actually supergiants (rank 1).  We then normalize these to the
number of stars in the lowest mass bin, 12-15$M_\odot$.  What we {\it observe} is a decreasing number of
yellow supergiants as we go up in mass.  At the bottom of Table~\ref{tab:numbers}
 we have included an ``extra" mass bin, 15-25$M_\odot$, as it is clear
from Table~\ref{tab:ages} that the evolutionary tracks predict a much longer lifetime for the 20$M_\odot$ model than for
either the 15$M_\odot$ or the 25$M_\odot$, and we wanted to see what the agreement would look like if we ignored
this track.

We can estimate the number of yellow supergiants we expect from the models if we assume ``steady state" star formation in M31.
By this, we require only that the star formation rate  {\it averaged over the entire disk} has stayed about the same for the past 20 Myr.
In that case, the number of stars $N$
in a particular evolutionary phase within a mass bin extending from one mass ($m_1$) to another ($m_2$) 
will just be:
$$N_{m_1}^{m_2} = [m^\Gamma]_{m_1}^{m_2} \times \bar \tau$$
where $\Gamma$ is the slope of the initial mass function (taken here to be $-1.35$ following
Salpeter 1955; see also Massey 1998b), and $\bar \tau$ is the average duration of the evolutionary phase for masses $m_1$ and $m_2$.  In the final column of Table~\ref{tab:numbers}
we give the expected number according to the ``S3" models.   We have normalized the expected numbers relative to that of 12-15$M_\odot$.

Here we find relatively poor agreement.  According to the models, the evolutionary time scale for the yellow supergiant
phase increases significantly with mass (Table~\ref{tab:ages}), more than compensating for the loss of stars due to the mass function.
Thus we expect to find $\sim 9\times$ more yellow supergiants between the 15-20$M_\odot$ tracks than between the 12-15$M_\odot$, 
or $\sim 6\times$ more
between the 20-25$M_\odot$ tracks than between the 12-15$M_\odot$ tracks.  But, in reality, we find, 0.7-0.8$\times$ and 0.2$\times$ as
many, respectively.  (Note that our results are insensitive to whether
we count ``all" of our candidates or just the certain ones.)   Mostly this comes about because of the very long time predicted for the 20$M_\odot$
yellow supergiant stage  relative to that of the lower masses (78,300 years vs 5,300 years).  If we ignore the 20$M_\odot$ track we find
only slightly better agreement, as the number of stars observed in the 15-25$M_\odot$ is about the same as in the 12-15$M_\odot$ track,
while the models predict  $4\times$ as many.   In addition, the models predict 6$\times$ as many 25-40$M_\odot$ yellow supergiants
as those of 12-15$M_\odot$.  Based on this we expect to find 110-150 yellow supergiants with masses of 25-40$M_\odot$,
but instead we observe {\it none}.  We can tell from
Table~\ref{tab:numbers} that the problem would be even worse if we had used the predictions of the ``Old" $z=0.040$ tracks, as
the lifetimes are even longer for stars with masses $\ge$25$M_\odot$.  

We do note that the 40$M_\odot$ just barely enters the yellow supergiant realm.  The long duration of this time stage reflects the fact
that the star takes some time to turn around there.  We also recall that in our sample we included stars that had bluer colors than our definition of yellow supergiants.  So, if there were an
abundance of such stars we would expect to have observed them, and, yet, none show up in the HRD.
Nevertheless, if we were to assume that  the duration of the 40$M_\odot$ yellow supergiant
phase was 0 years, rather than the  50,800 years we've included, then the  number of expected stars between 25 and 40$M_\odot$ would be 1.5$\times$ that of the 12-15$M_\odot$ yellow
supergiants. Thus rather than the 110-150 between the 25 and 40$M_\odot$ we expect by using the 40$M_\odot$ lifetime, we would only expect 30-60 stars between 25 and 40$M_\odot$. However, this still results in a significant discrepancy with observations, as we observe
{\it no} yellow supergiants in this mass range.

We emphasize that even though we expect some minimal contamination of our sample by foreground objects,  the {\it maximum} contamination 
for the rank 1 objects (``mostly certain") is 15\%.   However, we obtain essentially the same ratios in Table~\ref{tab:numbers} whether we
count all of our candidates (rank 1 and rank 2) or just the rank 1 stars.  So, it appears that our conclusion is robust.

Recall also from Figure~\ref{fig:magcomp} that if anything our radial velocity survey was slightly biased
towards the brighter stars than the fainter.   We have made no allowance for this in the observed
ratios in Table~\ref{tab:numbers}, but to do so would {\it increase} the discrepancy.  The typical
12-15$M_\odot$ stars have $M_V=-7$, or $V=17.8$, while the 20-25$M_\odot$ stars have
$M_V=-9$, or $V=15.8$.  The lower mass stars are therefore under represented by perhaps a factor of $\sim 1.5$.  Thus the 0.2 nominal
observed ratio of the 20-25$M_\odot$ stars relative to the 12-15$M_\odot$ stars should actually be lower by a factor of 1.5, suggesting
that the disagreement with the 5.6 ratio predicted by the models is about a factor of 40.

Is this a problem with the higher mass evolutionary tracks predicting too long a time scale for the yellow supergiant stage, or with the
lower mass tracks predicting too short a time scale?  We can answer this indirectly by computing their expected lifetimes based upon
the relative number of yellow supergiants and unevolved (O-type) stars observed.  Using the LGGS data, Massey (2009) estimates the  number of unevolved massive stars with
masses $>20M_\odot$ 
is about 24,800 in M31.  The IMF-weighted H-burning lifetime is of order 5~Myr, and, assuming a constant star formation rate, we would
thus expect to see $5\times10^{-3}$ massive stars born each year.  We observe 8 total (certain and probable) yellow supergiants
above 20$M_\odot$.  Recall that our sample contains only 68\% of the stars located in the region for which we expect radial velocities to be $<-150$ km s$^{-1}$, and that region covered 73\% of the area of the entire LGGS, from which the number of unevolved massive
stars were estimated.    We expect then that the true number of yellow supergiants
with masses$> 20M_\odot$ is about 16.  Therefore, we can estimate the actual ages of the yellow supergiant stage
as 16/24800 $\times$ 5 Myr. This is about 3200 years, which is consistent with the life times the 12-15$M_\odot$ models predict,
but is at variance with the much longer time scales predicted by the models for higher mass yellow supergiants.  We suggest that
these are too long by more than an order of magnitude.

Could this discrepancy instead be an argument that the global star formation rate in M31 has in fact not been constant over the past
20~Myr?  Yellow supergiants of 12$M_\odot$ are roughly 17~Myr old, according to the models, while those of 25$M_\odot$ are only 
7~Myr.  So, if during that 10~Myr period the overall star formation rate had decreased by a factor of 30-40 that would roughly
compensate for the smaller number of stars that we find.  However, such a drastic change is
in conflict with other observations.  Williams (2003) analyzed LGGS photometry and concluded
that there has been a slight (25\%) increase in the star formatation rate since 25~Myr ago.  This is a
marginal result, and consistent with constant star formation to within 2$\sigma$ (see his Table 2), 
but it certainly precludes the possibility of a 3000\% - 4000\% decrease over a similar time span.

\section{\label{fut} Summary, Discussion,  and and Future Work}

We measured radial velocities for $\sim 2900$ stars in M31, identifying 54 as rank 1 (nearly certain supergiants) and 66 as rank 2 (probable
supergiants).  Follow up spectroscopy eliminated 4 of the rank 2 stars, while confirming others as supergiants.  The magnitude limits
we chose should make our sample complete down to 12$M_\odot$.  In all we observed 68\% of the target candidates and the sample was
restricted to the 73\% of the LGGS area that should have radial velocities $<-150$ km s$^{-1}$. So, the true number of yellow supergiants
should be about a factor of 2.0 larger than what we find.  The foreground contamination proved to be 96-98\%.  There may be a few 
halo yellow giants among our candidates, but comparison with the Besancon model suggests this should be minor, at most 15\% for the
rank 1 objects based purely upon the distribution of radial velocities.  In practice we expect this contamination to be considerably less, since
we replied upon the difference between the observed and expected velocities (where the latter is dependent upon position in the galaxy)
to assign rank and membership.  Nevertheless, it would be very useful to conduct follow up spectroscopy of the rank 2 objects in order
to ascertain which have strong O I $\lambda 7774$ absorption or other spectral indicators of high luminosity.

We compared the location and numbers of yellow supergiants in the H-R diagram to those expected from the Geneva evolutionary
tracks.  We find excellent agreement between the locations of stars in the H-R diagram and the tracks: there are not (for instance)
high luminosity yellow supergiants with moderate temperatures that are inconsistent with the tracks.   Rather, the inconsistencies we do note
are related to the lifetimes predicted by the models for the yellow supergiant stage.  The number of yellow supergiants
decrease with increasing luminosity (mass), with no stars found more massive 25$M_\odot$.  Yet, the long duration of the yellow
supergiant phase predicted by the models for 20-40$M_\odot$ suggests that we should see far more high luminosity yellow supergiants
than what we observe.  Comparing the number of yellow supergiants we find to the number of unevolved O-type stars, suggests that
the typical duration of the yellow supergiant stage for stars with masses $>20M_\odot$ should be 3000 years, similar to what the
models predict for 12-15$M_\odot$ Yet, the models 
predict lifetimes far greater than this.   

We do not yet have an adequate explanation for the discrepancy, but will address this in
future work.   The higher mass tracks (for which the predicted yellow supergiant lifetimes appear to be too long) show the
stars evolving back to the blue after the RSG stage.   If instead the stars ended their lives as RSGs, without this loop back
to the blue, then the predicted yellow supergiant phase would be shorter as the star would pass through this region only once.
This might be the case if the mass loss during the RSG stage had been significantly over-estimated.  In part, this could be
tested by comparing the number of observed Wolf-Rayet stars (WRs) with that of RSGs.  The number of WRs in M31 is not known well enough
to make this comparison as yet.  Alternatively, it could be that the blue loops are present, but that the mass-loss rates during the evolution
blue-wards have been underestimated.  A sensitive test would be to conduct abundance studies of the yellow supergiants in this
region of the HRD to look for evidence that any of these stars are in a post-RSG phase.  An additional test
would be to look for circumstellar material left from the slow dense wind of the RSG stage around any of these objects.

It would, of course, be of interest to extend this work to galaxies with other metallicities, such as the Magellanic Clouds, where
the unevolved massive star content is also known, and where the models predict long lifetimes for yellow supergiants even for 12-15$M_\odot$
stars (Table~\ref{tab:ages}).  We hope to carry out such work ourselves during the next observing season.

\acknowledgements
We gratefully acknowledge the fine support at the MMT Observatory.  M.R.D.'s work was supported through
a National Science Foundation REU grant, AST-0453611, while P.M.'s efforts were supported through AST-0604569.
We are grateful to the Kitt Peak National Observatory Director, Buell Jannuzi, and to Di Harmer for their efforts in obtaining
the follow-up spectra; Harmer and Brian Skiff also offered useful advice that aided in our interpretation of these data.  Knut Olsen made useful comments on the recent star formation history of M31, and Heather Morrison provided useful correspondence on the issue of foreground
halo stars.  An anonymous referee provided thoughtful and constructive comments which helped improve the paper.

\clearpage
\begin{deluxetable}{l r r r r r r r r r r r}
\tablecaption{\label{tab:ages}Theoretical Yellow Supergiant Duration (yr)\tablenotemark{a}}
\tablewidth{0pt}
\tablehead{
&\multicolumn{3}{c}{Solar Neighbor.}
&
&\multicolumn{3}{c}{SMC}
&
&\multicolumn{3}{c}{M31} \\
\colhead{Mass}&\multicolumn{3}{c}{$z=0.020$}
&
&\multicolumn{3}{c}{$z=0.004$}
&
&\multicolumn{3}{c}{$z=0.040$} \\ \cline{2-4} \cline{6-8} \cline{10-12}
\colhead{($M_\odot$)}
&\colhead{S3\tablenotemark{b}}
&\colhead{S0\tablenotemark{b}}
&\colhead{Old\tablenotemark{c}}
&
&\colhead{S3\tablenotemark{d}}
&\colhead{S0\tablenotemark{e}}
&\colhead{Old\tablenotemark{f}}
&
&\colhead{S3\tablenotemark{d}}
&\colhead{S0\tablenotemark{d}}
&\colhead{Old\tablenotemark{g}}
}
\startdata
85 & 0 &  0  & 0                  &&  \nodata & \nodata & 12300  &&  0 & \nodata & 0 \\
60 & 0 & 14100 & 5300                 &&  14000 & 56500 & 62300                 &&   0 & 0 & 5400 \\
40 & 57600 & 403600 & 198100          &&   148100 & 32700 & 188200             &&   50800 & \nodata & 87300\\
25 & 5400 & 73000 & 23600               &&  \nodata & \nodata & 109800&&    18500 & \nodata & 666000 \\
20 & 300 & 92500 & 64800          &&  71700 & 16500   &  58900              &&   78300 & \nodata & 3800  \\
15 & 2100 & 50800 & 2300                  &&    206600 &  60400  &  92000            &&   3200 & \nodata & 1900\\
12 & 3600 & 32800 & 51000                &&    33200 &  30700  &  21800             &&    5300 &  \nodata & 2500\\
\enddata
\tablenotetext{a}{For the purposes of this calculation, yellow supergiants are defined as having effective temperatures between 4800~K and 7500~K. Note that not all versions of the evolutionary models are available for each mass and metallicity.}
\tablenotetext{b}{S3 has initial rotation 300 km s$^{-1}$, and S0 has an initial rotation of 0 km s$^{-1}$; ages were determined using models from Meynet \& Maeder (2003).}
\tablenotetext{c}{Rotation not included; determined using models from Schaller et al.\ (1992).}
\tablenotetext{d}{S3 has initial rotation 300 km s$^{-1}$, and S0 has an initial rotation of 0 km s$^{-1}$; ages were determined using models from Meynet \& Maeder (2005), except for the 9, 12, 20 and 25 $M_\odot$, $z=0.040$ models newly computed for this study and that of Massey et al.\ (2009), and the 12, 15 and 20 $M_\odot$, $z=0.004$ models from Maeder \& Meynet (2001).}
\tablenotetext{e}{The S0 models for $z=0.004$ were computed with an initial rotation of 0 km s$^{-1}$; ages were determined using the models
from Maeder \& Meynet (2001).}
\tablenotetext{f}{Rotation not included; determined using models from Charbonnel et al.\ (1993).}
\tablenotetext{g}{Rotation not included; determined using models from Schaerer et al.\ (1993).}
\end{deluxetable}

\clearpage

\begin{deluxetable}{l c c c l c}
\tablecaption{\label{tab:fields} Configurations Observed}
\tablewidth{0pt}
\tablehead{
\colhead{Config}
&\colhead{$\alpha_{\rm 2000}$}
&\colhead{$\delta_{\rm 2000}$}
&\colhead{Exps}
&\colhead{UT Date(s)}
&\colhead{$N$ stars observed} 
}
\startdata
Brt1-1 & 00 46 49.0&  +42 11 21  &   3x10min &  2007 10 23 &15\\
Brt2-1 & 00 44 00.8&  +41 37 28 &    3x10min &  2007 10 14  &92\\
Brt3-1 & 00 41 39.6 & +40 51 30 &    3x10min &  2007 10 14  &93\\
Brt4-1 & 00 40 46.9 & +40 36 36 &    3x10min &  2007 10 16  &111\\
Brt5-1 & 00 39 32.4&  +40 21 33 &    3x10min &  2007 10 17, 2007 10 23  &128\\
Fnt1-1 & 00 38 55.8 & +40 07 38  &   3x15min  & 2007 10 14  &201\\
Fnt1-2 & 00 39 09.8&  +40 09 24  &   3x15min  & 2007 10 14  &202\\
Fnt1-3 & 00 38 48.1 & +40 06 49 &    3x15min & 2007 10 16  &191\\
Fnt1-4 & 00 38 56.0&  +40 07 37 &    3x15min &  2007 10 16  &170\\
Fnt1-5&  00 39 06.3&  +40 06 54 &    3x15min &  2007 10 17  &148\\
Fnt1-6 & 00 39 09.0 & +40 07 06 &    3x15min &  2007 10 18  &109\\
Fnt2-1 & 00 41 48.9 & +40 55 23 &    3x15min &  2007 10 19  &199\\
Fnt2-2 & 00 41 37.9 & +40 54 30 &    3x15min &  2007 10 19  &198\\
Fnt2-3 & 00 41 26.7&  +40 54 49 &    3x15min &  2007 10 20  &200\\
Fnt2-4&  00 41 26.8&  +40 54 13 &    3x15min &  2007 10 20  &191\\
Fnt2-5&  00 41 38.0 & +40 54 30 &    3x15min &  2007 10 21  &150\\
Fnt2-6&  00 41 50.8 & +40 55 17 &    3x15min &  2007 10 21  &137\\
Fnt3-1 & 00 44 05.6&  +41 34 16 &    3x15min &  2007 11 20  &151\\
Fnt3-2&  00 44 29.3&  +41 38 49  &   3x15min &  2007 11 20  &150\\
Fnt3-3 & 00 44 36.2&  +41 43 33 &    3x15min  & 2007 11 20  &151\\
\enddata
\end{deluxetable}

\begin{deluxetable}{l c c r r r r c c l }
\tabletypesize{\tiny}
\tablecaption{\label{tab:all} Stars with Observed Radial Velocities\tablenotemark{*}}
\tablewidth{0pt}
\tablehead{
\colhead{Star}
&\colhead{$\alpha_{\rm 2000}$}
&\colhead{$\delta_{\rm 2000}$}
&\multicolumn{1}{r}{Vel$_{\rm obs}$}
&\multicolumn{1}{r}{$r$\tablenotemark{a}}
&\multicolumn{1}{r}{Vel$_{\rm expect}$}
&\multicolumn{1}{r}{Vel$_{\rm obs}$$-$Vel$_{\rm exp}$}
&\colhead{$V$}
&\colhead{$B-V$}
&\colhead{Rank\tablenotemark{b}} \\
& & &
\multicolumn{1}{r}{km s$^{-1}$}
&
&\multicolumn{1}{r}{km s$^{-1}$}
&\multicolumn{1}{r}{km s$^{-1}$}
}
\startdata
J003702.13+400945.6     &00:37:02.127    &+40:09:45.53    &   14.8    &  19.8& -513.5    &  528.3    &  16.09    &   1.29 &3   \\
J003702.47+401742.5     &00:37:02.467    &+40:17:42.42    &  -26.4    &  21.3& -487.7    &  461.3    &  18.03    &   1.32 &3   \\
J003702.80+400516.8     &00:37:02.797    &+40:05:16.74    &   -4.5    &  32.1& -524.9    &  520.4    &  16.79    &   0.71 &3   \\
J003702.94+400027.2     &00:37:02.937    &+40:00:27.14    & -122.8    &  32.6& -532.8    &  410.0    &  17.96    &   0.84 &3   \\
J003703.78+395541.6     &00:37:03.777    &+39:55:41.55    &  -42.0    &  41.4& -536.3    &  494.3    &  15.76    &   0.74 &3   \\
J003703.85+401402.8     &00:37:03.847    &+40:14:02.73    &  -34.1    &  25.1& -501.2    &  467.1    &  17.38    &   1.22 &3   \\
J003703.99+395939.5     &00:37:03.987    &+39:59:39.44    &  -47.9    &  30.7& -533.9    &  486.0    &  17.08    &   0.69 &3   \\
J003704.12+401702.0     &00:37:04.117    &+40:17:01.93    &   -6.6    &  24.3& -491.1    &  484.5    &  15.91    &   0.97 &3   \\
J003704.53+401426.0     &00:37:04.527    &+40:14:25.93    &  -48.7    &  19.6& -500.4    &  451.7    &  17.88    &   0.60 &3   \\
J003704.56+400521.0     &00:37:04.557    &+40:05:20.94    &  -50.5    &  36.0& -525.5    &  475.0    &  17.83    &   0.77 &3   \\
\enddata
\tablenotetext{*}{The full version of this table is available in the on-line edition.}
\tablenotetext{a}{Tonry \& Davis 1979 $r$ parameter.}
\tablenotetext{b}{Rank: 1=highly likely supergiant; 2=probable supergiant; 3=dwarf; Cl=cluster}
\tablenotetext{c}{Radial velocity observed on two different nights differed by 10 km s$^{-1}$ or more.}
\end{deluxetable}

\begin{deluxetable}{l c c r r r r c c l }
\tabletypesize{\tiny}
\tablecaption{\label{tab:observed} Observed Properties of Probable M31 Members}
\tablewidth{0pt}
\tablehead{
\colhead{Star}
&\colhead{$\alpha_{\rm 2000}$}
&\colhead{$\delta_{\rm 2000}$}
&\multicolumn{1}{r}{Vel$_{\rm obs}$}
&\multicolumn{1}{r}{$r$\tablenotemark{a}}
&\multicolumn{1}{r}{Vel$_{\rm expect}$}
&\multicolumn{1}{r}{Vel$_{\rm obs}$$-$Vel$_{\rm exp}$}
&\colhead{$V$}
&\colhead{$B-V$}
&\colhead{Rank\tablenotemark{b}} \\
& & &
\multicolumn{1}{r}{km s$^{-1}$}
&
&\multicolumn{1}{r}{km s$^{-1}$}
&\multicolumn{1}{r}{km s$^{-1}$}
}
\startdata
J003745.26+395823.6     &00:37:45.257    &+39:58:23.54    & -521.7    &  20.0& -533.4    &   11.7    &  17.16    &   0.59 &1   \\
J003907.59+402628.4     &00:39:07.587    &+40:26:28.32    & -533.6    &  18.9& -535.1    &    1.5    &  16.74    &   0.86 &1   \\
J003922.08+402031.5     &00:39:22.077    &+40:20:31.42    & -504.2    &  10.0& -527.8    &   23.6    &  18.36    &   0.18 &1   \\
J003926.72+402239.4     &00:39:26.717    &+40:22:39.32    & -541.7    &   8.2& -529.6    &  -12.1    &  18.07    &   0.08 &1   \\
J003930.55+403135.2     &00:39:30.547    &+40:31:35.11    & -547.1    &  30.3& -535.2    &  -11.9    &  18.15    &   0.78 &1   \\
J003930.79+401841.1     &00:39:30.787    &+40:18:41.02    & -518.0    &  19.2& -516.6    &   -1.4    &  16.90    &   0.46 &1   \\
J003935.23+401947.7     &00:39:35.227    &+40:19:47.62    & -515.9    &  20.8& -516.0    &    0.1    &  17.92    &   0.98 &1   \\
J003943.43+403524.9     &00:39:43.427    &+40:35:24.81    & -501.6    &  16.5& -533.8    &   32.2    &  17.17    &   0.23 &1   \\
J003948.28+403856.4     &00:39:48.277    &+40:38:56.30    & -519.9    &  35.4& -527.0    &    7.1    &  17.88    &   0.92 &1   \\
J003948.85+403844.8     &00:39:48.847    &+40:38:44.70    & -509.9    &  25.3& -528.1    &   18.2    &  17.40    &   0.65 &1   \\
J003949.86+405305.8     &00:39:49.857    &+40:53:05.69    & -440.3    &  24.2& -438.6    &   -1.7    &  17.34    &   1.02 &1   \\
J003953.55+402827.7     &00:39:53.547    &+40:28:27.61    & -539.5    &  35.1& -525.9    &  -13.6    &  17.45    &   1.18 &1   \\
J004009.13+403142.9     &00:40:09.127    &+40:31:42.81    & -554.8    &  25.5& -522.4    &  -32.4    &  17.63    &   0.84 &1   \\
J004009.70+403719.0     &00:40:09.697    &+40:37:18.91    & -554.4    &   9.3& -535.7    &  -18.7    &  18.27    &   0.20 &1   \\
J004017.72+400436.6     &00:40:17.717    &+40:04:36.54    & -409.1    &  13.6& -439.4    &   30.3    &  18.33    &   0.26 &1   \\
J004020.37+403428.8     &00:40:20.367    &+40:34:28.71    & -554.2    &  13.4& -520.8    &  -33.4    &  18.43    &   0.24 &1   \\
J004021.21+403117.1     &00:40:21.207    &+40:31:17.01    & -527.0    &  16.6& -507.5    &  -19.5    &  16.65    &   0.28 &1   \\
J004021.64+403256.5     &00:40:21.637    &+40:32:56.41    & -525.2    &  22.9& -513.5    &  -11.7    &  17.83    &   0.51 &1   \\
J004025.48+405041.0     &00:40:25.477    &+40:50:40.89    & -469.2    &  11.7& -501.4    &   32.2    &  17.99    &   0.13 &1   \\
J004026.68+403604.4     &00:40:26.677    &+40:36:04.31    & -552.4    &   7.7& -519.9    &  -32.5    &  18.44    &   0.24 &1   \\
J004029.38+403604.2     &00:40:29.377    &+40:36:04.11    & -562.5    &  17.5& -516.7    &  -45.8    &  18.13    &   0.31 &1   \\
J004030.10+402943.1     &00:40:30.097    &+40:29:43.01    & -502.8\tablenotemark{c}    &   8.6& -489.9    &  -12.9    &  18.27    &   0.15 &1   \\
J004030.62+404523.8     &00:40:30.617    &+40:45:23.70    & -562.3    &  18.5& -535.2    &  -27.1    &  18.33    &   0.29 &1   \\
J004032.90+404352.8     &00:40:32.897    &+40:43:52.70    & -576.6    &  32.5& -536.4    &  -40.2    &  18.12    &   0.92 &1   \\
J004034.00+405358.3     &00:40:33.997    &+40:53:58.19    & -500.9    &  20.0& -485.6    &  -15.3    &  18.02    &   0.46 &1   \\
J004034.00+403500.1     &00:40:33.997    &+40:35:00.01    & -540.0    &  19.6& -506.3    &  -33.7    &  17.70    &   0.44 &1   \\
J004035.37+405701.0     &00:40:35.367    &+40:57:00.89    & -406.2    &   7.2& -458.2    &   52.0    &  18.11    &   0.20 &1   \\
J004038.10+403827.2     &00:40:38.097    &+40:38:27.10    & -554.5    &  23.7& -515.9    &  -38.6    &  17.59    &   0.96 &1   \\
J004053.97+403256.2     &00:40:53.967    &+40:32:56.11    & -468.0    &  14.2& -469.1    &    1.1    &  18.31    &   0.80 &1   \\
J004101.24+410434.6     &00:41:01.237    &+41:04:34.48    & -427.7    &   2.4& -416.0    &  -11.7    &  17.12    &   0.92 &1   \\
J004101.76+410429.2     &00:41:01.757    &+41:04:29.08    & -422.4    &   6.7& -417.5    &   -4.9    &  17.42    &   0.82 &1   \\
J004102.78+410900.6     &00:41:02.777    &+41:09:00.47    & -362.0    &  21.2& -378.1    &   16.1    &  18.35    &   0.55 &1   \\
J004118.69+403152.0     &00:41:18.687    &+40:31:51.91    & -501.5    &   8.6& -433.3    &  -68.2    &  18.14    &   0.67 &1   \\
J004120.56+403515.4     &00:41:20.557    &+40:35:15.31    & -432.9    &  22.1& -439.5    &    6.6    &  17.10    &   0.60 &1   \\
J004120.99+403453.5     &00:41:20.987    &+40:34:53.41    & -526.1    &  21.8& -437.9    &  -88.2    &  18.13    &   0.45 &1   \\
J004128.74+405224.7     &00:41:28.737    &+40:52:24.59    & -519.3    &  17.9& -507.5    &  -11.8    &  17.79    &   0.25 &1   \\
J004129.31+405102.9     &00:41:29.307    &+40:51:02.79    & -511.5    &   2.2& -496.1    &  -15.4    &  16.87    &   0.72 &1   \\
J004131.50+403917.8     &00:41:31.497    &+40:39:17.70    & -403.1    &  17.2& -435.6    &   32.5    &  17.96    &   0.56 &1   \\
J004143.45+403956.4     &00:41:43.447    &+40:39:56.30    & -458.6    &  21.4& -420.6    &  -38.0    &  18.00    &   0.41 &1   \\
J004144.76+402808.9     &00:41:44.757    &+40:28:08.81    & -342.9    &  20.0& -401.1    &   58.2    &  18.29    &   1.03 &1   \\
J004149.87+412712.7     &00:41:49.867    &+41:27:12.56    & -313.0    &   6.6& -286.1    &  -26.9    &  17.99    &   0.83 &1   \\
J004156.90+412109.0     &00:41:56.897    &+41:21:08.86    & -334.8    &  19.2& -303.3    &  -31.5    &  17.17    &   0.54 &1   \\
J004201.09+403951.9     &00:42:01.087    &+40:39:51.80    & -424.2    &  16.1& -399.2    &  -25.0    &  18.46    &   0.36 &1   \\
J004207.22+405148.3     &00:42:07.217    &+40:51:48.19    & -415.8    &  27.0& -413.3    &   -2.5    &  17.02    &   0.97 &1   \\
J004207.85+405152.4     &00:42:07.847    &+40:51:52.29    & -414.8    &  18.5& -412.3    &   -2.5    &  16.99    &   0.76 &1   \\
J004212.20+405513.9     &00:42:12.197    &+40:55:13.79    & -453.0    &  16.4& -413.9    &  -39.1    &  18.24    &   0.23 &1   \\
J004214.85+405652.0     &00:42:14.847    &+40:56:51.89    & -457.0    &  15.2& -413.6    &  -43.4    &  17.97    &   0.19 &1   \\
J004215.06+405148.3     &00:42:15.057    &+40:51:48.19    & -381.5    &  10.1& -399.7    &   18.2    &  18.48    &   0.88 &1   \\
J004226.53+410123.9     &00:42:26.527    &+41:01:23.78    & -371.7    &  11.6& -400.0    &   28.3    &  18.37    &   0.47 &1   \\
J004229.30+405727.6     &00:42:29.297    &+40:57:27.49    & -412.1    &   9.8& -385.2    &  -26.9    &  17.63    &   0.43 &1   \\
J004233.76+410014.6     &00:42:33.757    &+41:00:14.48    & -389.6    &  24.7& -380.7    &   -8.9    &  18.03    &   0.68 &1   \\
J004247.25+410039.2     &00:42:47.247    &+41:00:39.08    & -370.0    &   9.9& -358.0    &  -12.0    &  18.21    &   0.80 &1   \\
J004259.95+410220.3     &00:42:59.947    &+41:02:20.18    & -423.3    &   4.4& -340.2    &  -83.1    &  17.41    &   1.06 &1   \\
J004304.89+410345.9     &00:43:04.887    &+41:03:45.78    & -341.5    &  11.6& -332.5    &   -9.0    &  18.07    &   0.89 &1   \\
J003711.98+395445.2     &00:37:11.977    &+39:54:45.15    & -331.9    &  18.8& -536.3    &  204.4    &  18.06    &   0.56 &2   \\
J003725.57+400731.9     &00:37:25.567    &+40:07:31.83    & -435.2    &  10.4& -529.7    &   94.5    &  18.47    &   0.50 &2   \\
J003934.02+404714.2     &00:39:34.017    &+40:47:14.10    & -269.1    &  24.8& -463.0    &  193.9    &  17.21    &   0.60 &2   \\
J003936.96+400743.8     &00:39:36.957    &+40:07:43.73    & -343.0    &  11.8& -480.4    &  137.4    &  18.19    &   0.80 &2   \\
J003937.44+394941.1     &00:39:37.437    &+39:49:41.05    & -305.7    &  10.5& -445.6    &  139.9    &  17.50    &   0.46 &2   \\
J003942.35+404031.8     &00:39:42.347    &+40:40:31.70    & -313.9    &  13.0& -514.5    &  200.6    &  18.15    &   0.54 &2   \\
J003955.87+401636.4     &00:39:55.867    &+40:16:36.33    & -332.8    &  23.7& -486.0    &  153.2    &  18.43    &   0.54 &2   \\
J004002.91+400659.2     &00:40:02.907    &+40:06:59.13    & -268.9    &   9.8& -455.6    &  186.7    &  17.98    &   0.40 &2   \\
J004007.14+410321.8     &00:40:07.137    &+41:03:21.68    & -209.5    &  18.1& -387.1    &  177.6    &  18.08    &   0.54 &2   \\
J004020.37+410723.2     &00:40:20.367    &+41:07:23.08    & -287.6    &  15.1& -371.3    &   83.7    &  18.32    &   0.68 &2   \\
J004034.82+401825.5     &00:40:34.817    &+40:18:25.42    & -306.9    &  23.4& -450.4    &  143.5    &  16.30    &   0.61 &2   \\
J004107.40+405328.6     &00:41:07.397    &+40:53:28.49    & -402.7    &  14.2& -535.9    &  133.2    &  17.50    &   0.47 &2   \\
J004148.69+413814.1     &00:41:48.687    &+41:38:13.95    & -215.7    &  27.5& -267.8    &   52.1    &  18.49    &   0.63 &2   \\
J004217.15+403740.2     &00:42:17.147    &+40:37:40.11    & -268.7    &  36.1& -381.9    &  113.2    &  17.77    &   0.66 &2   \\
J004223.21+412803.2     &00:42:23.207    &+41:28:03.06    & -303.0    &  17.5& -259.3    &  -43.7    &  18.36    &   0.67 &2   \\
J004231.84+412039.9     &00:42:31.837    &+41:20:39.76    & -230.1    &  16.1& -270.4    &   40.3    &  18.44    &   0.55 &2   \\
J004232.90+412103.4     &00:42:32.897    &+41:21:03.26    & -217.4    &  27.2& -265.9    &   48.5    &  17.43    &   0.86 &2   \\
J004235.91+413050.6     &00:42:35.907    &+41:30:50.45    & -265.0    &  11.8& -240.3    &  -24.7    &  17.80    &   0.47 &2   \\
J004237.52+413024.8     &00:42:37.517    &+41:30:24.65    & -261.4    &  11.9& -238.5    &  -22.9    &  17.33    &   0.24 &2   \\
J004240.34+412807.2     &00:42:40.337    &+41:28:07.06    & -248.9    &  15.1& -235.5    &  -13.4    &  18.01    &   0.52 &2   \\
J004243.54+414620.6     &00:42:43.537    &+41:46:20.44    & -176.1    &  34.8& -228.5    &   52.4    &  17.20    &   0.74 &2   \\
J004245.97+414422.1     &00:42:45.967    &+41:44:21.94    & -237.3    &  12.9& -226.5    &  -10.8    &  17.99    &   0.18 &2   \\
J004247.30+414451.0     &00:42:47.297    &+41:44:50.84    & -246.1    &  17.6& -225.4    &  -20.7    &  16.41    &   0.47 &2   \\
J004248.11+403434.5     &00:42:48.107    &+40:34:34.41    & -220.8    &  28.5& -360.1    &  139.3    &  17.20    &   0.76 &2   \\
J004250.67+414008.6     &00:42:50.667    &+41:40:08.44    & -259.1    &  27.9& -221.3    &  -37.8    &  17.85    &   0.67 &2   \\
J004251.90+413745.9     &00:42:51.897    &+41:37:45.75    & -210.1    &  30.5& -218.9    &    8.8    &  14.98    &   0.62 &2   \\
J004252.87+412328.6     &00:42:52.867    &+41:23:28.46    & -181.4    &  13.9& -192.5    &   11.1    &  18.30    &   0.09 &2   \\
J004255.16+413515.3     &00:42:55.157    &+41:35:15.15    & -245.6    &  19.1& -212.7    &  -32.9    &  17.84    &   0.83 &2   \\
J004259.62+413946.1     &00:42:59.617    &+41:39:45.94    & -364.6    &  12.7& -210.2    & -154.4    &  18.48    &   0.49 &2   \\
J004301.96+405315.2     &00:43:01.957    &+40:53:15.09    & -240.3    &  31.0& -346.2    &  105.9    &  17.78    &   0.79 &2   \\
J004302.55+413332.0     &00:43:02.547    &+41:33:31.85    & -138.3    &  52.1& -196.6    &   58.3    &  17.87    &   0.75 &2   \\
J004303.69+414543.3     &00:43:03.687    &+41:45:43.14    & -225.5    &  13.1& -209.9    &  -15.6    &  18.21    &   0.22 &2   \\
J004311.34+414240.9     &00:43:11.337    &+41:42:40.74    & -158.0    &  31.1& -197.7    &   39.7    &  18.17    &   0.79 &2   \\
J004313.02+414144.9     &00:43:13.017    &+41:41:44.74    & -178.1    &  35.5& -193.8    &   15.7    &  16.97    &   0.63 &2   \\
J004314.47+414229.1     &00:43:14.467    &+41:42:28.94    & -199.5    &  10.7& -192.9    &   -6.6    &  18.13    &   0.10 &2   \\
J004318.57+415311.1     &00:43:18.567    &+41:53:10.93    & -143.7    &  59.9& -201.0    &   57.3    &  16.54    &   0.78 &2   \\
J004325.93+413910.6     &00:43:25.927    &+41:39:10.45    & -182.0    &  48.2& -163.6    &  -18.4    &  18.32    &   0.75 &2   \\
J004337.16+412151.0     &00:43:37.157    &+41:21:50.86    & -221.6    &  23.5& -178.8    &  -42.8    &  17.02    &   0.91 &2   \\
J004338.76+414915.1     &00:43:38.757    &+41:49:14.94    & -197.5    &   8.5& -171.5    &  -26.0    &  18.16    &   0.26 &2   \\
J004348.01+415406.2     &00:43:48.007    &+41:54:06.03    & -148.1    &  31.4& -169.9    &   21.8    &  18.30    &   0.59 &2   \\
J004406.32+420131.3     &00:44:06.317    &+42:01:31.12    & -188.3    &  19.2& -163.5    &  -24.8    &  15.60    &   0.46 &2   \\
J004409.23+415941.1     &00:44:09.227    &+41:59:40.93    & -102.0    &  37.1& -156.1    &   54.1    &  17.42    &   0.62 &2   \\
J004409.98+420121.1     &00:44:09.977    &+42:01:20.92    & -145.1    &  23.5& -159.0    &   13.9    &  17.42    &   0.60 &2   \\
J004410.62+411759.7     &00:44:10.617    &+41:17:59.57    & -244.5    &  21.1& -243.3    &   -1.2    &  17.63    &   0.40 &2   \\
Mag-253496                     &00:44:12.450    &+41:16:08.42    & -226.9    &  54.5& -253.6    &   26.7    &  15.15    &   1.18 &2   \\
J004424.21+412116.0     &00:44:24.207    &+41:21:15.86    & -238.0    &  14.3& -225.9    &  -12.1    &  16.73    &   0.91 &2   \\
J004427.76+412209.8     &00:44:27.757    &+41:22:09.66    & -213.9    &  19.9& -221.3    &    7.4    &  17.26    &   1.03 &2   \\
J004428.99+412010.7     &00:44:28.987    &+41:20:10.56    & -216.8    &   9.1& -233.9    &   17.1    &  17.80    &   0.31 &2   \\
J004432.01+412442.0     &00:44:32.007    &+41:24:41.86    & -168.2    &   5.4& -205.3    &   37.1    &  18.28    &   0.07 &2   \\
J004432.41+412947.5     &00:44:32.407    &+41:29:47.35    & -120.2    &  59.4& -159.3    &   39.1    &  17.05    &   0.73 &2   \\
J004440.60+412704.1     &00:44:40.597    &+41:27:03.96    & -189.8    &  27.8& -192.7    &    2.9    &  18.22    &   1.33 &2   \\
J004441.56+412636.6     &00:44:41.557    &+41:26:36.46    & -165.4    &   9.3& -196.9    &   31.5    &  17.96    &   0.44 &2   \\
J004444.50+412314.3     &00:44:44.497    &+41:23:14.16    & -186.3    &  40.6& -220.8    &   34.5    &  17.57    &   0.65 &2   \\
J004447.45+412409.7     &00:44:47.447    &+41:24:09.56    & -197.9    &  12.3& -216.4    &   18.5    &  18.10    &   0.22 &2   \\
J004458.01+413217.5     &00:44:58.007    &+41:32:17.35    & -128.8    &  11.6& -165.7    &   36.9    &  17.79    &   0.17 &2   \\
J004508.90+413117.8     &00:45:08.897    &+41:31:17.65    & -144.0    &  32.6& -182.5    &   38.5    &  17.12    &   0.61 &2   \\
J004509.86+413031.5     &00:45:09.857    &+41:30:31.35    & -402.6    &   6.3& -188.3    & -214.3    &  16.83    &   0.56 &2   \\
J004518.17+413615.6     &00:45:18.167    &+41:36:15.45    & -101.1    &  41.2& -154.4    &   53.3    &  17.84    &   0.68 &2   \\
J004518.76+413630.7     &00:45:18.757    &+41:36:30.55    & -128.4    &  19.5& -153.1    &   24.7    &  16.70    &   0.51 &2   \\
J004526.93+412613.6     &00:45:26.927    &+41:26:13.46    & -255.0    &  24.6& -218.8    &  -36.2    &  18.45    &   0.75 &2   \\
J004532.62+413227.8     &00:45:32.617    &+41:32:27.65    & -436.7    &  35.3& -190.2    & -246.5    &  15.79    &   0.85 &2   \\
J004535.23+413600.5     &00:45:35.227    &+41:36:00.35    & -154.3    &  75.8& -170.8    &   16.5    &  15.88    &   0.94 &2   \\
J004554.48+413359.8     &00:45:54.477    &+41:33:59.65    & -344.4    &  20.1& -193.1    & -151.3    &  17.16    &   0.59 &2   \\
J004559.84+414038.2     &00:45:59.837    &+41:40:38.04    & -161.9    &   5.9& -161.0    &   -0.9    &  17.71    &   0.22 &2   \\
J004618.59+414410.9     &00:46:18.587    &+41:44:10.74    & -149.2    &  34.8& -154.6    &    5.4    &  15.22    &   0.65 &2   \\
J004658.64+414948.4     &00:46:58.637    &+41:49:48.24    & -134.6    &  15.0& -153.1    &   18.5    &  18.30    &   0.48 &2   \\
Mag-237751              &00:41:01.181    &+41:13:45.83    & -441.5    &  41.1& -348.0    &  -93.5    &  14.13    &   1.12 &Cl  \\
J004345.23+410608.5     &00:43:45.227    &+41:06:08.38    & -402.1    &  28.3& -298.2    & -103.9    &  18.32    &   0.52 &Cl  \\
J004356.46+412203.3     &00:43:56.457    &+41:22:03.16    & -362.0    &  59.1& -203.1    & -158.9    &  16.88    &   0.08 &Cl  \\
J004358.15+412438.8     &00:43:58.147    &+41:24:38.66    & -380.4    &  47.9& -171.4    & -209.0    &  17.35    &   0.99 &Cl  \\
J004403.98+412618.7     &00:44:03.977    &+41:26:18.56    & -293.1    &  26.5& -157.6    & -135.5    &  18.17    &   0.71 &Cl  \\
J004446.42+412918.3     &00:44:46.417    &+41:29:18.16    & -209.8    &  63.5& -179.4    &  -30.4    &  17.96    &   0.26 &Cl  \\
J004545.58+413942.4     &00:45:45.577    &+41:39:42.25    & -204.9    &  48.5& -155.5    &  -49.4    &  18.13    &   0.89 &Cl  \\
\enddata
\tablenotetext{a}{Tonry \& Davis 1979 $r$ parameter.}
\tablenotetext{b}{Rank: 1=highly likely supergiant; 2=probable supergiant; 3=dwarf; Cl=cluster}
\tablenotetext{c}{Radial velocity observed on two different nights differed by 10 km s$^{-1}$ or more.}
\end{deluxetable}

\begin{deluxetable}{l c c c c l}
\tabletypesize{\tiny}
\tablecaption{\label{tab:derived} Derived Properties of Potential M31 Supergiants}
\tablewidth{0pt}
\tablehead{
\colhead{Star}
&\colhead{Rank}
&\colhead{$M_V$}
&\colhead{$T_{eff}$}
&\colhead{$\log{(L/L_{\sun})}$}
&\colhead{Comment}
}
\startdata
Mag-253496              &2    &  (-9.65)& (3.713)    & (5.87) & No O I $\lambda 7774$--Nonmember\\
J004251.90+413745.9     &2    &  (-9.82)& (3.792)    & (5.84) & No O I $\lambda 7774$--Nonmember\\
J004618.59+414410.9     &2    &  (-9.58)& (3.787)    & (5.75) & Non-member\\
J004406.32+420131.3     &2    &  -9.20& 3.822    & 5.58\\
J004532.62+413227.8     &2    &  -9.01& 3.758    & 5.55\\
J004535.23+413600.5     &2    &  -8.92& 3.746    & 5.53 & Strong O I $\lambda 7774$\\
J004034.82+401825.5     &2    &  -8.50& 3.794    & 5.31\\
J004247.30+414451.0     &2    &  -8.39& 3.821    & 5.25 & Very strong O I $\lambda 7774$\\
J004318.57+415311.1     &2    &  -8.26& 3.768    & 5.24\\
J004424.21+412116.0     &2    &  -8.07& 3.750    & 5.18 & Strong O I $\lambda 7774$\\
J003907.59+402628.4     &1    &  -8.06& 3.757    & 5.17\\
J004021.21+403117.1     &1    &  -8.15& 3.867    & 5.15\\
J004518.76+413630.7     &2    &  -8.10& 3.812    & 5.14 \\
J004129.31+405102.9     &1    &  -7.93& 3.776    & 5.10\\
J004509.86+413031.5     &2    &  (-7.97)& (3.804)    & (5.09) & Weak O I $\lambda 7774$--Non member?\\
J004207.22+405148.3     &1    &  -7.77& 3.743    & 5.08\\
J004337.16+412151.0     &2    &  -7.78& 3.750    & 5.07 & Strong O I $\lambda 7774$\\
J004207.85+405152.4     &1    &  -7.81& 3.771    & 5.06\\
J003930.79+401841.1     &1    &  -7.90& 3.823    & 5.05\\
J004313.02+414144.9     &2    &  -7.83& 3.791    & 5.04\\
J004101.24+410434.6     &1    &  -7.68& 3.749    & 5.03\\
J004432.41+412947.5     &2    &  -7.75& 3.776    & 5.03\\
J004120.56+403515.4     &1    &  -7.70& 3.796    & 4.99\\
J004427.76+412209.8     &2    &  -7.54& 3.734    & 4.99\\
J004508.90+413117.8     &2    &  -7.68& 3.794    & 4.98\\
J004248.11+403434.5     &2    &  -7.59& 3.771    & 4.97\\
J004243.54+414620.6     &2    &  -7.60& 3.774    & 4.97\\
J003949.86+405305.8     &1    &  -7.46& 3.736    & 4.96\\
J004554.48+413359.8     &2    &  -7.64& 3.797    & 4.96\\
J004156.90+412109.0     &1    &  -7.63& 3.807    & 4.96 & Very strong O I $\lambda 7774$\\
J003745.26+395823.6     &1    &  -7.64& 3.797    & 4.96\\
J003943.43+403524.9     &1    &  -7.63& 3.879    & 4.95\\
J003953.55+402827.7     &1    &  -7.35& 3.712    & 4.95\\
J003934.02+404714.2     &2    &  -7.59& 3.795    & 4.94\\
J004259.95+410220.3     &1    &  -7.39& 3.730    & 4.94\\
J004101.76+410429.2     &1    &  -7.38& 3.762    & 4.89\\
J004237.52+413024.8     &2    &  -7.47& 3.877    & 4.89\\
J004232.90+412103.4     &2    &  -7.37& 3.756    & 4.89\\
J003948.85+403844.8     &1    &  -7.40& 3.787    & 4.88\\
J004409.98+420121.1     &2    &  -7.38& 3.796    & 4.86\\
J004409.23+415941.1     &2    &  -7.38& 3.793    & 4.86\\
J004038.10+403827.2     &1    &  -7.21& 3.743    & 4.85\\
J003937.44+394941.1     &2    &  -7.30& 3.823    & 4.82\\
J004107.40+405328.6     &2    &  -7.30& 3.821    & 4.81\\
J004444.50+412314.3     &2    &  -7.23& 3.787    & 4.81\\
J004009.13+403142.9     &1    &  -7.17& 3.760    & 4.81\\
J004410.62+411759.7     &2    &  -7.17& 3.835    & 4.76\\
J004229.30+405727.6     &1    &  -7.17& 3.830    & 4.76\\
J004301.96+405315.2     &2    &  -7.02& 3.767    & 4.74\\
J004559.84+414038.2     &2    &  -7.09& 3.884    & 4.74\\
J004217.15+403740.2     &2    &  -7.03& 3.787    & 4.73\\
J004255.16+413515.3     &2    &  -6.96& 3.761    & 4.73\\
J004034.00+403500.1     &1    &  -7.10& 3.827    & 4.73\\
J003935.23+401947.7     &1    &  -6.88& 3.741    & 4.72\\
J003948.28+403856.4     &1    &  -6.92& 3.750    & 4.72\\
J004458.01+413217.5     &2    &  -7.01& 3.898    & 4.71\\
J004440.60+412704.1     &2    &  -6.58& 3.688    & 4.70\\
J004250.67+414008.6     &2    &  -6.95& 3.785    & 4.70\\
J004302.55+413332.0     &2    &  -6.93& 3.773    & 4.70\\
J004518.17+413615.6     &2    &  -6.95& 3.783    & 4.70\\
J004128.74+405224.7     &1    &  -7.01& 3.876    & 4.70 & Very strong O I $\lambda 7774$\\
J004428.99+412010.7     &2    &  -7.00& 3.857    & 4.69\\
J004021.64+403256.5     &1    &  -6.97& 3.813    & 4.69\\
J004235.91+413050.6     &2    &  -7.00& 3.820    & 4.69\\
J003926.72+402239.4     &1    &  -6.73& 3.961    & 4.67\\
J004149.87+412712.7     &1    &  -6.81& 3.761    & 4.66\\
J004025.48+405041.0     &1    &  -6.81& 3.932    & 4.66\\
J004214.85+405652.0     &1    &  -6.83& 3.891    & 4.64\\
J004131.50+403917.8     &1    &  -6.84& 3.803    & 4.64\\
J004304.89+410345.9     &1    &  -6.73& 3.753    & 4.64\\
J004441.56+412636.6     &2    &  -6.84& 3.827    & 4.63\\
J004245.97+414422.1     &2    &  -6.81& 3.896    & 4.63 & Strong O I $\lambda 7774$\\
J004314.47+414229.1     &2    &  -6.67& 3.951    & 4.63\\
J004032.90+404352.8     &1    &  -6.68& 3.749    & 4.63\\
J004233.76+410014.6     &1    &  -6.77& 3.783    & 4.63 & Very strong O I $\lambda 7774$\\
J004002.91+400659.2     &2    &  -6.82& 3.836    & 4.62\\
J004240.34+412807.2     &2    &  -6.79& 3.810    & 4.62\\
J004034.00+405358.3     &1    &  -6.78& 3.823    & 4.61 & Strong O I $\lambda 7774$\\
J004143.45+403956.4     &1    &  -6.80& 3.833    & 4.61\\
J003711.98+395445.2     &2    &  -6.74& 3.803    & 4.60\\
J003930.55+403135.2     &1    &  -6.65& 3.769    & 4.59\\
J004432.01+412442.0     &2    &  -6.52& 3.966    & 4.59\\
J004311.34+414240.9     &2    &  -6.63& 3.767    & 4.59\\
J004007.14+410321.8     &2    &  -6.72& 3.807    & 4.59\\
J004035.37+405701.0     &1    &  -6.69& 3.888    & 4.58\\
J003936.96+400743.8     &2    &  -6.61& 3.765    & 4.58\\
J004144.76+402808.9     &1    &  -6.51& 3.734    & 4.58\\
J004118.69+403152.0     &1    &  -6.66& 3.784    & 4.58\\
J004447.45+412409.7     &2    &  -6.70& 3.882    & 4.58\\
J004247.25+410039.2     &1    &  -6.59& 3.765    & 4.57\\
J004252.87+412328.6     &2    &  -6.50& 3.954    & 4.57\\
J004029.38+403604.2     &1    &  -6.67& 3.857    & 4.56\\
J003942.35+404031.8     &2    &  -6.65& 3.806    & 4.56\\
J004120.99+403453.5     &1    &  -6.67& 3.825    & 4.56\\
J004338.76+414915.1     &2    &  -6.64& 3.871    & 4.55\\
J004303.69+414543.3     &2    &  -6.59& 3.884    & 4.54\\
J004030.10+402943.1     &1    &  -6.53& 3.924    & 4.54\\
J004053.97+403256.2     &1    &  -6.49& 3.766    & 4.53\\
J004325.93+413910.6     &2    &  -6.48& 3.773    & 4.52\\
J004212.20+405513.9     &1    &  -6.56& 3.881    & 4.52\\
J004020.37+410723.2     &2    &  -6.48& 3.783    & 4.51\\
J004348.01+415406.2     &2    &  -6.50& 3.797    & 4.51\\
J004009.70+403719.0     &1    &  -6.53& 3.889    & 4.51\\
J004658.64+414948.4     &2    &  -6.50& 3.819    & 4.50\\
J004223.21+412803.2     &2    &  -6.43& 3.785    & 4.49\\
J004030.62+404523.8     &1    &  -6.47& 3.864    & 4.48 & Strong O I $\lambda 7774$\\
J004017.72+400436.6     &1    &  -6.47& 3.873    & 4.48\\
J004102.78+410900.6     &1    &  -6.45& 3.806    & 4.48\\
J004215.06+405148.3     &1    &  -6.32& 3.754    & 4.48\\
J003922.08+402031.5     &1    &  -6.44& 3.897    & 4.48\\
J004226.53+410123.9     &1    &  -6.43& 3.820    & 4.47\\
J004526.93+412613.6     &2    &  -6.35& 3.772    & 4.47\\
J004231.84+412039.9     &2    &  -6.36& 3.804    & 4.45\\
J003955.87+401636.4     &2    &  -6.37& 3.807    & 4.45\\
J004026.68+403604.4     &1    &  -6.36& 3.878    & 4.44\\
J004020.37+403428.8     &1    &  -6.37& 3.876    & 4.44\\
J004148.69+413814.1     &2    &  -6.31& 3.790    & 4.44\\
J004201.09+403951.9     &1    &  -6.34& 3.845    & 4.43 & Strong O I $\lambda 7774$\\
J004259.62+413946.1     &2    &  -6.32& 3.816    & 4.43\\
J003725.57+400731.9     &2    &  -6.33& 3.815    & 4.43\\
\enddata
\end{deluxetable}

\begin{deluxetable}{l r r r r r}
\tabletypesize{\small}
\tablecaption{\label{tab:numbers} Number of Yellow Supergiants Compared to Models}
\tablewidth{0pt}
\tablehead{
\colhead{Mass}
&\colhead{\#}
&\colhead{\#}
&\multicolumn{3}{c}{Ratio relative to 12-15$M_\odot$}  \\ \cline{4-6}
\colhead{Range}
&\colhead{All}
&\colhead{Certain}
&\colhead{All}
&\colhead{Certain}
&\colhead{Models}
}
\startdata
12-15$M_\odot$   & 41  & 20 & 1.0 & 1.0 & 1.0 \\
15-20$M_\odot$   & 28  & 16 & 0.7 & 0.8 & 8.7 \\
20-25$M_\odot$   & 8    &   4  &0.2 & 0.2 & 5.7 \\
25-40$M_\odot$   & 0   &  0   &0.0 & 0.0 & 5.5 \\
15-25$M_\odot$   & 36  & 20 & 0.9 & 1.0 & 3.6 \\

\enddata
\end{deluxetable}
\end{document}